\documentclass[amsmath,amssymb,nofootinbib]{revtex4}
\usepackage{axodraw4j}
\usepackage{pstricks}
\usepackage{color}
\usepackage{graphicx}
\usepackage{dcolumn}
\usepackage{bm}

\begin{document}

\title{NLO corrections for the dipole factorization\\
 of DIS structure functions at low $x$}
\author{Guillaume Beuf}
\email{guillaume.beuf@usc.es}
\affiliation{Department of Physics,
Brookhaven National Laboratory,\\
Upton, NY 11973, USA\\}
\affiliation{Departamento de F\'isica de Part\'iculas and IGFAE, Universidade de Santiago de
Compostela,\\
 E-15706 Santiago de Compostela, Spain\\}

\begin{abstract}
The NLO generalization of the dipole factorization formula for the
structure functions $F_2$ and $F_L$ at low $x$ is calculated using light front perturbation theory.
That result gives some interesting insight into the kinematics of initial state parton showers in mixed space.
\end{abstract}

\pacs{13.60.Hb}

\maketitle


\section{Introduction}

QCD at low Bjorken $x$ has become a very active topic of research in particular thanks to deep inelastic scattering (DIS) experiments at HERA. A large part of the DIS phenomenology at low $x$ since the start of HERA has been based on the dipole factorization derived by Nikolaev and Zakharov in Ref.\cite{Nikolaev:1990ja}. Contrary to earlier works in low $x$ QCD, which have been performed in momentum space like the derivation of the BFKL evolution \cite{Lipatov:1976zz,Kuraev:1977fs,Balitsky:1978ic} resummming the high-energy leading logs (LL), the dipole factorization is formulated in mixed space, specifying the transverse position and the light-cone momentum of the involved particles.  The dipole factorization has been proposed not only for DIS structure functions \cite{Nikolaev:1990ja} but also for other DIS observables at low $x$, like diffractive structure functions \cite{Nikolaev:1991et}, deeply virtual Compton scattering and exclusive vector meson production \cite{Kopeliovich:1991pu}. The dipole factorization provides a very intuitive picture of DIS observables: the virtual photon radiated from the lepton first fluctuates into a quark-antiquark dipole, which then interacts with the target via gluon exchange(s). In the leading order (LO) version of the dipole factorization, the splitting probability for the photon into a dipole is known from perturbative QED. By contrast, the other factor, which is the dipole-target elastic scattering amplitude, contains all the QCD dynamics, both perturbative and non-perturbative.

Numerous phenomenological studies based on the dipole factorization have been performed. In most of them, phenomenological models for the dipole-target amplitude encoding various effects have been used. However, it has been soon realized that the BFKL evolution can be rederived as the low $x$ evolution of a dipole cascade in mixed space \cite{Mueller:1993rr,Mueller:1994jq}. Hence it is very natural to combine the two results, and pick the dipole-target  amplitude among the solutions of the BFKL equation in mixed space. By adding such constraint from perturbative QCD, one reduces \emph{a priori} the needed amount of modeling, down to the choice of the initial condition for the BFKL evolution.

However, the BFKL evolution has some severe shortcomings like the violation of unitarity at high energy and the sensitivity to the non-perturbative infrared physics, especially if the coupling is running. The phenomenon of gluon saturation at high energy \cite{Gribov:1984tu,Mueller:1985wy} is both a consequence of those issues and a quasi-perfect solution\footnote{Gluon saturation solves the problem of infrared sensitivity at large enough rapidity and restores the unitarity of the dipole-target amplitude at fixed impact parameter. However, it is not enough \cite{GolecBiernat:2003ym,Berger:2010sh} to prevent the violation of the Froissart bound \cite{Froissart:1961ux}, after integration over the impact parameter.} to them. When taking gluon saturation into account, the BFKL equation is generalized into the B-JIMWLK equations, derived both from the high-energy operator product expansion of Wilson line operators \cite{Balitsky:1995ub} and from the Color Glass Condensate (CGC) effective theory \cite{Jalilian-Marian:1997jx,Jalilian-Marian:1997gr,Jalilian-Marian:1997dw,Kovner:2000pt,Weigert:2000gi,Iancu:2000hn,Iancu:2001ad,Ferreiro:2001qy}
based on earlier works \cite{McLerran:1993ni,McLerran:1993ka,McLerran:1994vd}. In a mean-field approximation, the B-JIMWLK equations reduce to the Balitsky-Kovchegov (BK) equation, also derived \cite{Kovchegov:1999yj,Kovchegov:1999ua} independently in the framework of Mueller's dipole cascade \cite{Mueller:1993rr,Mueller:1994jq}.
It has been soon realized that when using the LO dipole factorization together with the LL BK or B-JIMWLK equations, one describes the DIS data only qualitatively, because the obtained low $x$ evolution is faster than in the data. One is thus led to consider higher order corrections.

As a first step towards the NLO/NLL accuracy, the first contributions to the running of the coupling $\alpha_s$ in the BK or B-JIMWLK equations have been calculated \cite{Balitsky:2006wa,Kovchegov:2006vj,Gardi:2006rp}, leading to appropriate prescriptions to set the scale of the running coupling in the BK and B-JIMWLK equations. By simply promoting the coupling the BK and B-JIMWLK equations to a running coupling following those prescriptions, an very good description of the low $x$ DIS data can already be achieved \cite{Albacete:2009fh,Albacete:2010sy,Kuokkanen:2011je}.

Then, the full NLL BK equation has been calculated \cite{Balitsky:2008zz,Balitsky:2009xg}. Finally, the NLO photon impact factor has been obtained \cite{Balitsky:2010ze} in full coordinate space in the high-energy operator expansion of the product of two electromagnetic currents. That coordinate space representation is especially convenient to study conformal symmetry breaking or restoration depending on the choice of factorization scheme for the LL resummation. That NLO calculation allows to check explicitly that the B-JIMWLK equations are properly resumming the LL. Notice that the ongoing effort towards NLO accuracy with gluon saturation effects is not limited to DIS observables. As an example, NLO corrections to inclusive forward hadron production in hadronic collisions  has been calculated most recently \cite{Chirilli:2011km}. That observable is probably the simplest one for hadronic collisions to be sensitive to gluon saturation. It also involves an effective gluon distribution directly related to the same dipole target amplitude as used for DIS observables.

Unfortunately, however, the results \cite{Balitsky:2010ze} for the NLO photon impact factor are not available in a form suitable for phenomenological studies. One needs to perform non-trivial Fourier transforms from full coordinate space to full momentum space for the incoming and outgoing photons in order to obtain the NLO generalization of the dipole factorization formula \cite{Nikolaev:1990ja}. The purpose of the present paper is to fill that gap. More exactly, the explicit Fourier transform of the previous results \cite{Balitsky:2010ze} is left for further studies, and instead the NLO generalization \eqref{sigma_TL_Y_fplus} of the dipole factorization formula for the virtual photon cross sections $\sigma^{\gamma}_{T}(x,Q^2)$
and $\sigma^{\gamma}_{L}(x,Q^2)$ is directly calculated. Let us remind that those cross sections are related to the DIS structure functions $F_L$ and $F_2\equiv F_T+F_L$ as
\begin{equation}
F_{T,L}(x,Q^2)=\frac{Q^2}{(2\pi)^2\, \alpha_{em}}\; \sigma^{\gamma}_{T,L}(x,Q^2)\, .
\end{equation}
The calculation presented here is based on the same general ideas as in ref.\cite{Balitsky:2010ze}, however the method differs significantly. Indeed, light-front perturbation theory \cite{Kogut:1969xa,Bjorken:1970ah} is used here whereas the more standard covariant perturbation theory is used in ref.\cite{Balitsky:2010ze}.

The results of both ref.\cite{Balitsky:2010ze} and the present paper should give, in the two-gluon exchange approximation, the appropriate NLO photon impact factor for the BFKL equation in mixed space. The NLO photon impact factor for the BFKL equation in momentum space has also been calculated previously \cite{Bartels:2001mv,Bartels:2002uz,Bartels:2004bi}, but is not available in a closed analytical form. The comparison
of the three approaches is left for further studies.

In the section \ref{sec:photonWaveFunctions}, the calculation of the quark-antiquark-gluon components of the light-front wave functions of transverse or longitudinal virtual photons is performed, leading to the results \eqref{qqbarglue_wavefunction}, \eqref{NLO_T_wavefunction} and \eqref{NLO_L_wavefunction}. The associated virtual corrections are inferred from probability conservation. Those results are buildings blocks not only for the NLO corrections to the structure functions, but
also to other DIS observables admitting a dipole factorization.

Those intermediate results also allows to shed some light on the kinematics of dipole cascades in mixed space.
A simple prescription \eqref{trans_recoil} to take recoil effects into account directly in mixed space is proposed. And a very simple expression in mixed space \eqref{form_time_conjecture} for the formation time of arbitrary multi-parton Fock states through initial state radiation is conjectured. That expression is purely kinematical and depends on the result of the parton cascade but not on the diagram followed. That expression should nevertheless contain full recoil effects diagram by diagram.

The results of the section \ref{sec:photonWaveFunctions} are used in the section \ref{sec:crossSection} to derive the NLO generalization \eqref{sigma_TL_Y_fplus} of the dipole factorization formula for the virtual photon cross sections $\sigma^{\gamma}_{L}(x,Q^2)$ and $\sigma^{\gamma}_{T}(x,Q^2)$. That expression involves the NLO impact factors \eqref{ImpFact_NLO_L} and \eqref{ImpFact_NLO_T}.
Conclusions are then given in the section \ref{sec:conclu}.

For completeness, a detailed description of the formalism and notations used for the calculation is provided in the appendix \ref{sec:formalism}.

\section{Virtual photon wave-functions\label{sec:photonWaveFunctions}}

As explained in appendix \ref{sec:formalism}, the first step in the calculation of the DIS cross section is to obtain the light-front wave-functions of longitudinal and transverse virtual photons at the appropriate order. Notice that if the photon stays a photon or splits into a lepton pair, it cannot interact with gluons of the target and thus does not give a large contribution to the DIS cross section at low $x$. So, we will always consider only the components of the photon wave-functions containing colored partons, and drop the other ones.

\subsection{Quark-antiquark components at LO}

\begin{figure}
\setbox1\hbox to 10cm{
\fcolorbox{white}{white}{
 \begin{picture}(592,252) (9,-9)
    \SetWidth{1.4}
    \SetColor{Black}
    \Photon(30,132)(170,132){7.5}{7}
    \Arc(240,132)(70,90,270)
    \Line[arrow,arrowpos=0.5,arrowlength=17,arrowwidth=6.8,arrowinset=0.4](240,202)(370,202)
    \Line[arrow,arrowpos=0.5,arrowlength=17,arrowwidth=6.8,arrowinset=0.4](370,62)(240,62)
    \Text(10,102)[lb]{\LARGE{\Black{$q^+, Q, \lambda$}}}
    \Text(380,192)[lb]{\LARGE{\Black{$\mathbf{k}_0 \textrm{ or } \mathbf{x}_0,\, k^+_0,\, h_0,\, A_0,\, f$}}}
    \Text(380,52)[lb]{\LARGE{\Black{$\mathbf{k}_1 \textrm{ or } \mathbf{x}_1,\, k^+_1,\, h_1,\, A_1,\, f$}}}
    \SetWidth{1.0}
    \SetColor{White}
    \EBox(10,-8)(600,242)
  \end{picture}
}
}
\begin{center}
\hspace{-5cm}\resizebox*{5cm}{!}{\box1}
\caption{\label{Fig:DiagLOT}LO diagram for the transverse photon wave function.}
\end{center}
\end{figure}
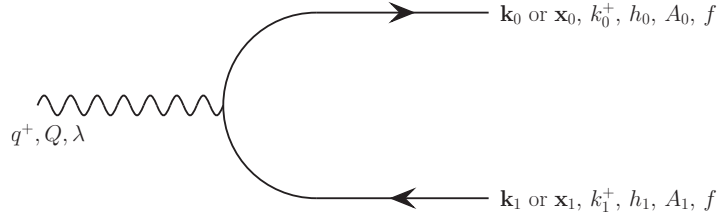

As a warm-up, let us outline how to obtain the standard LO results for the virtual photon wave-functions in the formalism based on light-front perturbation theory \cite{Kogut:1969xa,Bjorken:1970ah} described in appendix \ref{sec:formalism}. At this order, the only diagram contributing to the colored sector of the transverse photon wave-function is shown on Fig.\ref{Fig:DiagLOT}. For a photon of helicity $\lambda=\pm 1$, virtuality $Q$, momentum $q^+$ large enough and $\mathbf{q}=0$ one gets from the general formula \eqref{LFWF_pert} the incoming state in the Heisenberg picture
\begin{eqnarray}
\left|{\gamma_T^*\big(q^+,Q^2,\lambda\big)}_H\right\rangle_{LO} &=&\!\!\!\! \sum_{q\bar{q}\textrm{ states}} \!\!\!\! b^\dag(\mathbf{k}_0,k^+_0,h_0,A_0,f)  d^\dag(\mathbf{k}_1,k^+_1,h_1,A_1,f)  |0\rangle\: \frac{1}{\left(-\frac{Q^2}{2q^+}\!-\!k_0^-\!-\!k_1^-\!+\!i\epsilon\right)}\nonumber\\
& & \qquad \times
\langle 0 | d(\mathbf{k}_1,k^+_1,h_1,A_1,f)\,  b(\mathbf{k}_0,k^+_0,h_0,A_0,f)\,  \hat{{\cal U}}_I(0)\,  {a_{\gamma}}^\dag(\mathbf{0},q^+,\lambda)|0\rangle\nonumber\\
&=&\frac{e}{2(2\pi)} \int_0^1 \frac{\textrm{d}z_0}{\sqrt{z_0}} \int_0^1 \frac{\textrm{d}z_1}{\sqrt{z_1}}\, \delta(z_0\!+\!z_1\!-\!1) \sum_{h_0} \left[ z_1\!-\!z_0-(2 h_0) \lambda\right]\,
 \int \frac{\textrm{d}^2\mathbf{k}_{0}}{(2\pi)^2} \frac{\varepsilon_{\lambda}\cdot \mathbf{k}_0}{(z_0 z_1 Q^2+{\mathbf{k}_0}^2)}\nonumber\\
 & &\quad \times \sum_f e_f \sum_{A_0} \: b^\dag(\mathbf{k}_0,k^+_0,h_0,A_0,f)  d^\dag(-\mathbf{k}_0,k^+_1,-h_0,A_0,f)  |0\rangle\, ,
\end{eqnarray}
where $z_i=k^+_i/q^+$ is the longitudinal momentum fraction with respect to the momentum $q^+$ of the virtual photon.

For the case of longitudinal photon, the LO diagram for the wave-function is essentially the same as on Fig.\ref{Fig:DiagLOT}, but with the effective vertex \eqref{gammaL2qqbarVertexEff}, so that
\begin{eqnarray}
\left|{\gamma_L^*\big(q^+,Q^2\big)}_H\right\rangle_{LO} &=&\!\!\!\! \sum_{q\bar{q}\textrm{ states}} \!\!\!\! b^\dag(\mathbf{k}_0,\dots)  d^\dag(\mathbf{k}_1,\dots)  |0\rangle\: \frac{1}{\left(-\frac{Q^2}{2q^+}\!-\!k_0^-\!-\!k_1^-\!+\!i\epsilon\right)}\:
{\cal V}_{\gamma^*_L(q^+,Q^2)\rightarrow q(\mathbf{k}_0,\dots)\; \bar{q}(\mathbf{k}_1,\dots)}\nonumber\\
&=&-\frac{e}{(2\pi)} \int_0^1 \frac{\textrm{d}z_0}{\sqrt{z_0}} \int_0^1 \frac{\textrm{d}z_1}{\sqrt{z_1}}\, \delta(z_0\!+\!z_1\!-\!1)
 \int \frac{\textrm{d}^2\mathbf{k}_{0}}{(2\pi)^2} \frac{z_0 z_1 Q}{(z_0 z_1 Q^2+{\mathbf{k}_0}^2)}\nonumber\\
 & &\quad \times \sum_f e_f \sum_{A_0} \sum_{h_0} \: b^\dag(\mathbf{k}_0,k^+_0,h_0,A_0,f)  d^\dag(-\mathbf{k}_0,k^+_1,-h_0,A_0,f)  |0\rangle\, .
\end{eqnarray}

Performing Fourier transforms \eqref{bdagFourier} thanks to the relations \eqref{K1_LO} and \eqref{K0_LO}, one obtains the mixed space representation of these wave-functions, which write
\begin{eqnarray}
\left|{\gamma_{T,L}^*\big(q^+,Q^2,(\lambda)\big)}_H\right\rangle_{LO} &=&\frac{e}{2} \int_0^1 \frac{\textrm{d}z_0}{\sqrt{z_0}} \int_0^1 \frac{\textrm{d}z_1}{\sqrt{z_1}}\, \delta(z_0\!+\!z_1\!-\!1)  \int \frac{\textrm{d}^2\mathbf{x}_{0}}{(2\pi)^2}
 \int \frac{\textrm{d}^2\mathbf{x}_{1}}{(2\pi)^2}  \sum_{h_0}  \: \Phi^{LO}_{T,L}\Big(\mathbf{x}_{0},\mathbf{x}_{1},z_0,z_1,(h_0),(\lambda)\Big)
\nonumber\\
 & &\quad \times \sum_f e_f \sum_{A_0} \: b^\dag(\mathbf{x}_0,k^+_0,h_0,A_0,f)  d^\dag(\mathbf{x}_1,k^+_1,-h_0,A_0,f)  |0\rangle\, ,\label{LO_wavefunction}
\end{eqnarray}
with
\begin{eqnarray}
\Phi^{LO}_{T}\Big(\mathbf{x}_{0},\mathbf{x}_{1},z_0,z_1,h_0,\lambda\Big) &=& i  \left[ z_1\!-\!z_0 -(2 h_0) \lambda\right]\,
\frac{\varepsilon_{\lambda}\cdot \mathbf{x}_{01}}{{x}_{01}^2}\,   Q\sqrt{z_0 z_1 {x}_{01}^2}\,\, \textrm{K}_1\!\left(Q\sqrt{z_0 z_1 {x}_{01}^2}\right)\label{LO_T_wavefunction}\\
\Phi^{LO}_{L}\Big(\mathbf{x}_{0},\mathbf{x}_{1},z_0,z_1\Big)  &=& -2 z_0\, z_1\, Q\,\, \textrm{K}_0\!\left(Q\sqrt{z_0 z_1 {x}_{01}^2}\right)\, ,\label{LO_L_wavefunction}
\end{eqnarray}
using the notations $\mathbf{x}_{pq}=\mathbf{x}_{p}-\mathbf{x}_{q}$ and ${x}_{pq}=|\mathbf{x}_{pq}|$. And $\textrm{K}_0(x)$ and $\textrm{K}_1(x)$ are modified Bessel functions of the second kind. Those LO results are of course not new, and are consistent with previous calculations in the literature like in ref.\cite{Dosch:1996ss}.

\subsection{Quark-antiquark-gluon components in the NLO wave-functions}

At NLO, the colored sector of the virtual photon wave-functions contains both quark-antiquark and quark-antiquark-gluon components, so that on can write
\begin{equation}
\left|{\gamma_{T,L}^*\big(q^+,Q^2,(\lambda)\big)}_H\right\rangle_{NLO}= \left|{\gamma_{T,L}^*\big(q^+,Q^2,(\lambda)\big)}_H\right\rangle_{q\bar{q}}
+\left|{\gamma_{T,L}^*\big(q^+,Q^2,(\lambda)\big)}_H\right\rangle_{q\bar{q}g}\, .\label{full_NLO_wavefunction}
\end{equation}

\begin{figure}
\setbox1\hbox to 10cm{
\fcolorbox{white}{white}{
  \begin{picture}(592,252) (9,-9)
    \SetWidth{1.4}
    \SetColor{Black}
    \Photon(30,132)(170,132){7.5}{7}
    \Arc(240,132)(70,90,270)
    \Line[arrow,arrowpos=0.5,arrowlength=17,arrowwidth=6.8,arrowinset=0.4](240,202)(370,202)
    \Line[arrow,arrowpos=0.5,arrowlength=17,arrowwidth=6.8,arrowinset=0.4](370,62)(240,62)
    \Text(10,102)[lb]{\LARGE{\Black{$q^+, Q, \lambda$}}}
    \Text(380,192)[lb]{\LARGE{\Black{$\mathbf{k}_0 \textrm{ or } \mathbf{x}_0,\, k^+_0,\, h_0,\, A_0,\, f$}}}
    \Text(380,52)[lb]{\LARGE{\Black{$\mathbf{k}_1 \textrm{ or } \mathbf{x}_1,\, k^+_1,\, h_1,\, A_1,\, f$}}}
    \SetWidth{1.0}
    \SetColor{White}
    \EBox(10,-8)(600,242)
    \SetWidth{1.4}
    \SetColor{Black}
    \GluonArc(351.923,239.692)(99.351,-157.704,-79.517){7.5}{11}
    \Text(380,132)[lb]{\LARGE{\Black{$\mathbf{k}_2 \textrm{ or } \mathbf{x}_2,\, k^+_2,\, \lambda_2,\, a$}}}
    \Text(240,12)[lb]{\Huge{\Black{$(a)$}}}
  \end{picture}
}
}
\setbox2\hbox to 10cm{
\fcolorbox{white}{white}{
  \begin{picture}(592,252) (9,-9)
    \SetWidth{1.4}
    \SetColor{Black}
    \Photon(30,132)(170,132){7.5}{7}
    \Arc(240,132)(70,90,270)
    \Line[arrow,arrowpos=0.5,arrowlength=17,arrowwidth=6.8,arrowinset=0.4](240,202)(370,202)
    \Line[arrow,arrowpos=0.5,arrowlength=17,arrowwidth=6.8,arrowinset=0.4](370,62)(240,62)
    \Text(10,102)[lb]{\LARGE{\Black{$q^+, Q, \lambda$}}}
    \Text(380,192)[lb]{\LARGE{\Black{$\mathbf{k}_0 \textrm{ or } \mathbf{x}_0,\, k^+_0,\, h_0,\, A_0,\, f$}}}
    \Text(380,52)[lb]{\LARGE{\Black{$\mathbf{k}_1 \textrm{ or } \mathbf{x}_1,\, k^+_1,\, h_1,\, A_1,\, f$}}}
    \SetWidth{1.0}
    \SetColor{White}
    \EBox(10,-8)(600,242)
    \SetWidth{1.4}
    \SetColor{Black}
    \GluonArc[clock](350,12)(111.803,153.435,79.695){7.5}{12}
    \Text(380,112)[lb]{\LARGE{\Black{$\mathbf{k}_2 \textrm{ or } \mathbf{x}_2,\, k^+_2,\, \lambda_2,\, a$}}}
    \Text(230,12)[lb]{\Huge{\Black{$(b)$}}}
  \end{picture}
}
}
\setbox3\hbox to 10cm{
\fcolorbox{white}{white}{
  \begin{picture}(592,252) (9,-9)
    \SetWidth{1.4}
    \SetColor{Black}
    \Photon(30,62)(230,62){7.5}{10}
    \Line[arrow,arrowpos=0.5,arrowlength=17,arrowwidth=6.8,arrowinset=0.4](230,202)(370,202)
    \Line[arrow,arrowpos=0.5,arrowlength=17,arrowwidth=6.8,arrowinset=0.4](370,62)(230,62)
    \Text(10,82)[lb]{\LARGE{\Black{$q^+, Q, \lambda$}}}
    \Text(380,192)[lb]{\LARGE{\Black{$\mathbf{k}_0 \textrm{ or } \mathbf{x}_0,\, k^+_0,\, h_0,\, A_0,\, f$}}}
    \Text(380,52)[lb]{\LARGE{\Black{$\mathbf{k}_1 \textrm{ or } \mathbf{x}_1,\, k^+_1,\, h_1,\, A_1,\, f$}}}
    \SetWidth{1.0}
    \SetColor{White}
    \EBox(10,-8)(600,242)
    \SetWidth{1.4}
    \SetColor{Black}
    \GluonArc(510,662)(538.516,-121.329,-105.068){7.5}{12}
    \Text(380,132)[lb]{\LARGE{\Black{$\mathbf{k}_2 \textrm{ or } \mathbf{x}_2,\, k^+_2,\, \lambda_2,\, a$}}}
    \Line(230,62)(230,202)
    \SetWidth{1.0}
    \Line(220,132)(240,132)
    \Text(230,12)[lb]{\Huge{\Black{$(c)$}}}
  \end{picture}
}
}
\setbox4\hbox to 10cm{
\fcolorbox{white}{white}{
  \begin{picture}(592,252) (9,-9)
    \SetWidth{1.4}
    \SetColor{Black}
    \Photon(30,202)(230,202){7.5}{10}
    \Line[arrow,arrowpos=0.5,arrowlength=17,arrowwidth=6.8,arrowinset=0.4](230,202)(370,202)
    \Line[arrow,arrowpos=0.5,arrowlength=17,arrowwidth=6.8,arrowinset=0.4](370,62)(230,62)
    \Text(10,162)[lb]{\LARGE{\Black{$q^+, Q, \lambda$}}}
    \Text(380,192)[lb]{\LARGE{\Black{$\mathbf{k}_0 \textrm{ or } \mathbf{x}_0,\, k^+_0,\, h_0,\, A_0,\, f$}}}
    \Text(380,52)[lb]{\LARGE{\Black{$\mathbf{k}_1 \textrm{ or } \mathbf{x}_1,\, k^+_1,\, h_1,\, A_1,\, f$}}}
    \SetWidth{1.0}
    \SetColor{White}
    \EBox(10,-8)(600,242)
    \SetWidth{1.4}
    \SetColor{Black}
    \GluonArc[clock](510,-398)(538.516,121.329,105.068){7.5}{12}
    \Text(380,112)[lb]{\LARGE{\Black{$\mathbf{k}_2 \textrm{ or } \mathbf{x}_2,\, k^+_2,\, \lambda_2,\, a$}}}
    \Line(230,62)(230,202)
    \SetWidth{1.0}
    \Line(220,132)(240,132)
    \Text(230,12)[lb]{\Huge{\Black{$(d)$}}}
  \end{picture}
}
}
\begin{center}
\resizebox*{10cm}{!}{\hspace{-8cm}\mbox{\box1 \hspace{12cm} \box2}}
\resizebox*{10cm}{!}{\hspace{-8cm}\mbox{\box3 \hspace{12cm} \box4}}
\caption{\label{Fig:DiagNLOT}Diagrams for the quark-antiquark-gluon component the transverse photon wave function at NLO. Each diagram is ordered along $x^+$, from $x^+\rightarrow -\infty$ on the left to $x^+=0$ on the right.}
\end{center}
\end{figure}
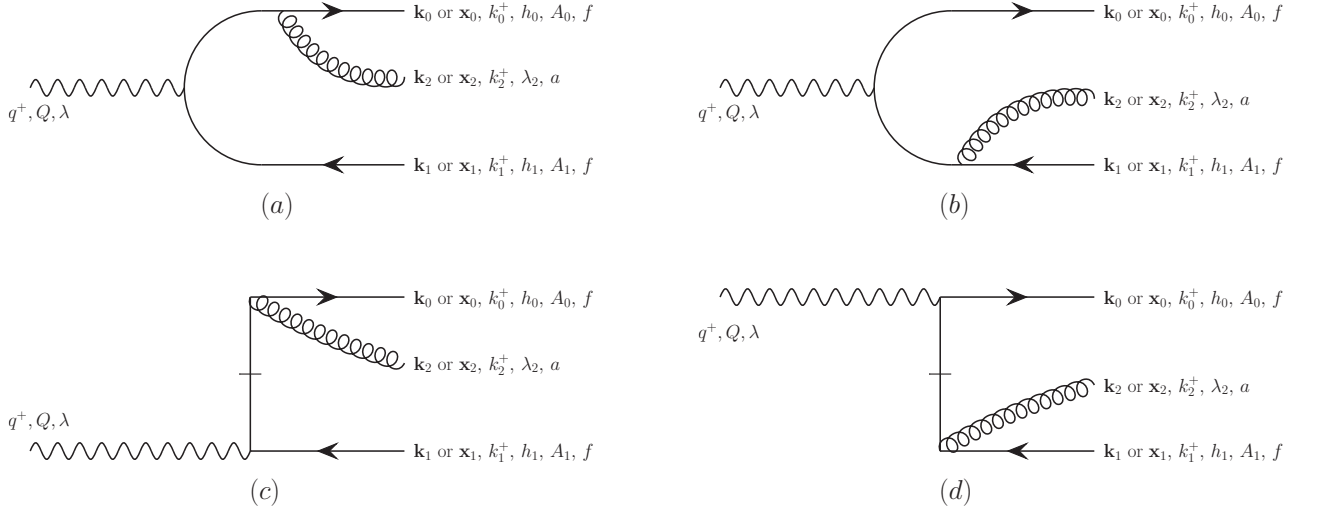
The four light-front diagrams on Fig.\ref{Fig:DiagNLOT}(a), \ref{Fig:DiagNLOT}(b), \ref{Fig:DiagNLOT}(c) and \ref{Fig:DiagNLOT}(d) contribute to the quark-antiquark-gluon component of the transverse photon wave-function at NLO. In the diagrams \ref{Fig:DiagNLOT}(a) and \ref{Fig:DiagNLOT}(b), the gluon is emitted from the quark or from the anti-quark at a later $x^+$ than the splitting of the photon into the quark-antiquark dipole. By contrast, in the diagrams \ref{Fig:DiagNLOT}(c) and \ref{Fig:DiagNLOT}(d),  the splitting of the photon into a quark-antiquark-gluon system is instantaneous in $x^+$. Those two different instantaneous interactions correspond to the two terms in the instantaneous vertex \eqref{inst_gamma2qqbarg_vertex}. Hence, one has
\begin{eqnarray}
\left|{\gamma_T^*\big(q^+,Q^2,\lambda\big)}_H\right\rangle_{q\bar{q}g} &=&\!\!\!\! \sum_{q\bar{q}g \textrm{ states}} \!\!\!\!  \frac{b^\dag(\mathbf{k}_0,\dots)  d^\dag(\mathbf{k}_1,\dots) a^\dag(\mathbf{k}_2,\dots)  |0\rangle}{\left(-\frac{Q^2}{2q^+}\!-\!k_0^-\!-\!k_1^-\!-\!k_2^-\!+\!i\epsilon\right)}\nonumber\\
& & \times \Bigg\{\sum_{\textrm{int. } q \textrm{ states}} \frac{ \langle 0 | a(\mathbf{k}_2,\dots)  b(\mathbf{k}_0,\dots)  \hat{{\cal U}}_I(0)  b^\dag(\mathbf{k}_i,\dots)|0\rangle\:
\langle 0 | d(\mathbf{k}_1,\dots)  b(\mathbf{k}_i,\dots)  \hat{{\cal U}}_I(0)  {a_{\gamma}}^\dag(\mathbf{0},q^+,\lambda)|0\rangle}{\left(-\frac{Q^2}{2q^+}\!-\!k_1^-\!-\!k_i^-\!+\!i\epsilon\right)}  \nonumber\\
& & \quad -\!\!\!\sum_{\textrm{int. } \bar{q} \textrm{ states}} \frac{ \langle 0 | a(\mathbf{k}_2,\dots)  d(\mathbf{k}_1,\dots)  \hat{{\cal U}}_I(0)  d^\dag(\mathbf{k}_i,\dots)|0\rangle\:
\langle 0 | d(\mathbf{k}_i,\dots)  b(\mathbf{k}_0,\dots)  \hat{{\cal U}}_I(0)  {a_{\gamma}}^\dag(\mathbf{0},q^+,\lambda)|0\rangle}{\left(-\frac{Q^2}{2q^+}\!-\!k_0^-\!-\!k_i^-\!+\!i\epsilon\right)}  \nonumber\\
& & \quad +
\langle 0 | a(\mathbf{k}_2,\dots) d(\mathbf{k}_1,\dots) b(\mathbf{k}_0,\dots)  \hat{{\cal U}}_I(0){a_{\gamma}}^\dag(\mathbf{0},q^+,\lambda) |0\rangle
\Bigg\}\, ,
\end{eqnarray}
where the second, third and fourth lines correspond respectively to the contributions of the diagrams \ref{Fig:DiagNLOT}(a), \ref{Fig:DiagNLOT}(b) and \ref{Fig:DiagNLOT}(c)+\ref{Fig:DiagNLOT}(d). For the diagram \ref{Fig:DiagNLOT}(a) (resp. \ref{Fig:DiagNLOT}(b)), there is a summation over the momentum $(k_i^+,\mathbf{k}_i)$ and quantum number of the quark (resp. antiquark) on the internal line.
Inserting the expressions \eqref{q2qg_vertex}, \eqref{qbar2qbarg_vertex}, \eqref{gamma2qqbar_vertex} and \eqref{inst_gamma2qqbarg_vertex} for the vertices and performing elementary algebra, one obtains the momentum space expression
\begin{eqnarray}
\left|{\gamma_T^*\big(q^+,Q^2,\lambda\big)}_H\right\rangle_{q\bar{q}g} &=& -\frac{e g}{2(2\pi)^2}  \sum_{h_0,\lambda_2}  \int_0^1 \frac{\textrm{d}z_0}{\sqrt{z_0}} \int_0^1 \frac{\textrm{d}z_1}{\sqrt{z_1}} \int_0^1 \frac{\textrm{d}z_2}{z_2}\, \delta(z_0\!+\!z_1\!+\!z_2\!-\!1) \: \int \frac{\textrm{d}^2\mathbf{k}_{0}}{(2\pi)^2} \int \frac{\textrm{d}^2\mathbf{k}_{1}}{(2\pi)^2} \int \frac{\textrm{d}^2\mathbf{k}_{2}}{(2\pi)^2}\nonumber\\
& &\times \Bigg\{
\frac{(1\!-\!z_1)}{(1\!-\!z_1\!-\!z_2)}
  \frac{\left[ 1\!-\!2 z_1\!+\!2 h_0\, \lambda\right]\, \left(\varepsilon_{\lambda}\!\cdot\! \mathbf{k}_{1}\right)}{\left(z_1(1\!-\!z_1)Q^2+{\mathbf{k}_{1}}^2\right)}  \left[ 1\!-\!\frac{z_2}{(1\!-\!z_1)}\left(\frac{1\!-\!2 h_0\, \lambda_2}{2}\right)\right]\, \varepsilon_{\lambda_2}^{*}\!\cdot\!
\left(\frac{\mathbf{k}_{2}}{z_2}\!+\!\frac{\mathbf{k}_{1}}{1\!-\!z_1} \right)\nonumber\\
& &\quad -\frac{(1\!-\!z_0)}{(1\!-\!z_0\!-\!z_2)}
  \frac{\left[ 1\!-\!2 z_0\!-\!2 h_0\, \lambda\right]\, \left(\varepsilon_{\lambda}\!\cdot\! \mathbf{k}_{0}\right)}{\left(z_0(1\!-\!z_0)Q^2+{\mathbf{k}_{0}}^2\right)} \left[1\!-\!\frac{z_2}{(1\!-\!z_0)}\left(\frac{1\!+\!2 h_0\, \lambda_2}{2}\right) \right]\, \varepsilon_{\lambda_2}^{*}\!\cdot\!
\left(\frac{\mathbf{k}_{2}}{z_2}\!+\!\frac{\mathbf{k}_{0}}{1\!-\!z_0} \right)\nonumber\\
& &\quad + \delta_{\lambda,\lambda_2}\, \left[\frac{ \delta_{\lambda,-2h_0}}{1\!-\!z_1}-\frac{ \delta_{\lambda,2h_0}}{1\!-\!z_0}\right]
\Bigg\}\quad
\frac{(2\pi)^2 \delta^{(2)}(\mathbf{k}_0\!+\!\mathbf{k}_1\!+\!\mathbf{k}_2)}{\left(Q^2+\frac{\mathbf{k}_0^2}{z_0}+\frac{\mathbf{k}_1^2}{z_1}
 +\frac{\mathbf{k}_2^2}{z_2}\right)}\nonumber\\
& &\times  \sum_f e_f\! \sum_{A_0,A_1,a}\! \big(T^a\big)_{A_0 A_1} b^\dag(\mathbf{k}_0,k^+_0,h_0,A_0,f)\,  d^\dag(\mathbf{k}_1,k^+_1,-h_0,A_1,f)\, a^\dag(\mathbf{k}_2,k^+_2,\lambda_2,a) |0\rangle
\, .\label{qqbarglue_T_wavefunction_mom}
\end{eqnarray}

In the case of a longitudinal photon, one needs to always start with the effective vertex \eqref{gammaL2qqbarVertexEff}. Since the latter already corresponds to a piece in a current-current instantaneous coulombian interaction, as explained in the appendix \ref{sec:lepton2photon}, there is no diagram with instantaneous splitting of a longitudinal photon into a quark-antiquark-gluon system. Hence, in the longitudinal photon case, one has only diagrams analog to \ref{Fig:DiagNLOT}(a) and \ref{Fig:DiagNLOT}(b), so that
\begin{eqnarray}
\left|{\gamma_L^*\big(q^+,Q^2\big)}_H\right\rangle_{q\bar{q}g} &=&\!\!\!\! \sum_{q\bar{q}g \textrm{ states}} \!\!\!\!  \frac{b^\dag(\mathbf{k}_0,\dots)  d^\dag(\mathbf{k}_1,\dots) a^\dag(\mathbf{k}_2,\dots)  |0\rangle}{\left(-\frac{Q^2}{2q^+}\!-\!k_0^-\!-\!k_1^-\!-\!k_2^-\!+\!i\epsilon\right)}\nonumber\\
& & \times \Bigg\{\sum_{\textrm{int. } q \textrm{ states}}\!\!\!\!\!\!\!\! \frac{ \langle 0 | a(\mathbf{k}_2,\dots)  b(\mathbf{k}_0,\dots)  \hat{{\cal U}}_I(0)  b^\dag(\mathbf{k}_i,\dots)|0\rangle\:
{\cal V}_{\gamma^*_L(q^+,Q^2)\rightarrow q(\mathbf{k}_i,\dots)\; \bar{q}(\mathbf{k}_1,\dots)}}{\left(-\frac{Q^2}{2q^+}\!-\!k_1^-\!-\!k_i^-\!+\!i\epsilon\right)}  \nonumber\\
& & -\!\!\!\!\!\!\!\!\!\!\sum_{\textrm{int. } \bar{q} \textrm{ states}}\!\!\!\!\!\!\!\! \frac{ \langle 0 | a(\mathbf{k}_2,\dots)  d(\mathbf{k}_1,\dots)  \hat{{\cal U}}_I(0)  d^\dag(\mathbf{k}_i,\dots)|0\rangle\:
{\cal V}_{\gamma^*_L(q^+,Q^2)\rightarrow q(\mathbf{k}_0,\dots)\; \bar{q}(\mathbf{k}_i,\dots)}}{\left(-\frac{Q^2}{2q^+}\!-\!k_0^-\!-\!k_i^-\!+\!i\epsilon\right)}
\Bigg\}\, ,
\end{eqnarray}
which gives, after substitution with the appropriate expressions,
\begin{eqnarray}
\left|{\gamma_L^*\big(q^+,Q^2\big)}_H\right\rangle_{q\bar{q}g} &=& \sum_{h_0,\lambda_2}  \int_0^1 \frac{\textrm{d}z_0}{\sqrt{z_0}} \int_0^1 \frac{\textrm{d}z_1}{\sqrt{z_1}} \int_0^1 \frac{\textrm{d}z_2}{z_2}\, \delta(z_0\!+\!z_1\!+\!z_2\!-\!1) \: \int \frac{\textrm{d}^2\mathbf{k}_{0}}{(2\pi)^2} \int \frac{\textrm{d}^2\mathbf{k}_{1}}{(2\pi)^2} \int \frac{\textrm{d}^2\mathbf{k}_{2}}{(2\pi)^2}\frac{(2\pi)^2 \delta^{(2)}(\mathbf{k}_0\!+\!\mathbf{k}_1\!+\!\mathbf{k}_2)}{\left(Q^2\!+\!\frac{\mathbf{k}_0^2}{z_0}\!
+\!\frac{\mathbf{k}_1^2}{z_1}\!+\!\frac{\mathbf{k}_2^2}{z_2}\right)}\nonumber\\
& &\times \Bigg\{
\frac{(1\!-\!z_1)^2 z_1}{(1\!-\!z_1\!-\!z_2)}\,
   \left[ 1\!-\!\frac{z_2}{(1\!-\!z_1)}\left(\frac{1\!-\!2 h_0\, \lambda_2}{2}\right)\right]\,
   \frac{\varepsilon_{\lambda_2}^{*}\!\cdot\!\left(\frac{\mathbf{k}_{2}}{z_2}\!+\!\frac{\mathbf{k}_{1}}{1\!-\!z_1} \right)}{\left(z_1(1\!-\!z_1)Q^2+{\mathbf{k}_{1}}^2\right)}
\nonumber\\
& &\quad -\frac{(1\!-\!z_0)^2 z_0}{(1\!-\!z_0\!-\!z_2)}\,
    \left[ 1\!-\!\frac{z_2}{(1\!-\!z_0)}\left(\frac{1\!+\!2 h_0\, \lambda_2}{2}\right)\right]\,
    \frac{\varepsilon_{\lambda_2}^{*}\!\cdot\! \left(\frac{\mathbf{k}_{2}}{z_2}\!+\!\frac{\mathbf{k}_{0}}{1\!-\!z_0} \right)}{\left(z_0(1\!-\!z_0)Q^2+{\mathbf{k}_{0}}^2\right)}
\Bigg\}\nonumber\\
& &\times  \frac{e g Q}{(2\pi)^2}  \:  \sum_f e_f\! \sum_{A_0,A_1,a}\! \big(T^a\big)_{A_0 A_1} b^\dag(\mathbf{k}_0,k^+_0,h_0,A_0,f)\,  d^\dag(\mathbf{k}_1,k^+_1,-h_0,A_1,f)\, a^\dag(\mathbf{k}_2,k^+_2,\lambda_2,a) |0\rangle
\, .\label{qqbarglue_L_wavefunction_mom}
\end{eqnarray}

The mixed space representation of those quark-antiquark-gluon components of the transverse and longitudinal photon wave-functions are obtained by Fourier transforming the creation operators as in \eqref{bdagFourier}. In both cases, one gets an expression of the type
\begin{eqnarray}
\left|{\gamma_{T,L}^*\big(q^+,Q^2,(\lambda)\big)}_H\right\rangle_{q\bar{q}g} &=&\frac{e g}{2(2\pi)} \int_0^1 \frac{\textrm{d}z_0}{\sqrt{z_0}} \int_0^1 \frac{\textrm{d}z_1}{\sqrt{z_1}} \int_0^1 \frac{\textrm{d}z_2}{z_2}\, \delta(z_0\!+\!z_1\!+\!z_2\!-\!1)  \int \frac{\textrm{d}^2\mathbf{x}_{0}}{(2\pi)^2}
 \int \frac{\textrm{d}^2\mathbf{x}_{1}}{(2\pi)^2}
  \int \frac{\textrm{d}^2\mathbf{x}_{2}}{(2\pi)^2}\nonumber\\
  & &\quad \times
 \sum_{h_0,\lambda_2}  \: \Phi^{q\bar{q}g}_{T,L}\Big(\mathbf{x}_{0},\mathbf{x}_{1},\mathbf{x}_{2},z_0,z_1,z_2,h_0,\lambda_2,(\lambda)\Big)\: \sum_f e_f \sum_{A_0,A_1,a}\: \big(T^a\big)_{A_0 A_1}
\nonumber\\
 & &\quad \times \: b^\dag(\mathbf{x}_0,k^+_0,h_0,A_0,f)\,  d^\dag(\mathbf{x}_1,k^+_1,-h_0,A_1,f)\, a^\dag(\mathbf{x}_2,k^+_2,\lambda_2,a) |0\rangle\, ,\label{qqbarglue_wavefunction}
\end{eqnarray}
where, using the integrals \eqref{K1_NLO_2steps}, \eqref{K1_NLO_inst} and \eqref{K0_NLO_2steps},
\begin{eqnarray}
& & \Phi^{q\bar{q}g}_{T}\Big(\mathbf{x}_{0},\mathbf{x}_{1},\mathbf{x}_{2},z_0,z_1,z_2,h_0,\lambda_2,\lambda\Big) = \frac{Q X\, \textrm{K}_1\!\left(Q X\right)}{X^2}\nonumber\\
& & \qquad \qquad \times\Bigg\{z_1(1\!-\!z_1)
  \left[ 1\!-\!2 z_1\!+\!2 h_0\, \lambda\right]\, \varepsilon_{\lambda}\!\cdot\! \left(\mathbf{x}_{10}\!-\!\frac{z_2}{1\!-\!z_1} \mathbf{x}_{20}\right)\, \left[ 1\!-\!\frac{z_2}{(1\!-\!z_1)}\left(\frac{1\!-\!2 h_0\, \lambda_2}{2}\right)\right]\,
\frac{\varepsilon_{\lambda_2}^{*}\!\cdot\!\mathbf{x}_{20}}{{x}_{20}^2}\nonumber\\
& &\qquad \qquad \quad -z_0(1\!-\!z_0) \left[1\!-\!2 z_0\!-\!2 h_0\, \lambda\right]\, \varepsilon_{\lambda}\!\cdot\! \left(\mathbf{x}_{01}\!-\!\frac{z_2}{1\!-\!z_0} \mathbf{x}_{21}\right)\, \left[ 1\!-\!\frac{z_2}{(1\!-\!z_0)}\left(\frac{1\!+\!2 h_0\, \lambda_2}{2}\right)\right]\,
\frac{ \varepsilon_{\lambda_2}^{*}\!\cdot\!\mathbf{x}_{21}}{{x}_{21}^2}\nonumber\\
& &\qquad \qquad \quad -z_0\, z_1\, z_2\, \delta_{\lambda,\lambda_2}\, \left[\frac{ \delta_{\lambda,-2h_0}}{1\!-\!z_1}-\frac{ \delta_{\lambda,2h_0}}{1\!-\!z_0}\right]
\Bigg\} \, . \label{NLO_T_wavefunction}
\end{eqnarray}
and
\begin{eqnarray}
\Phi^{q\bar{q}g}_{L}\Big(\mathbf{x}_{0},\mathbf{x}_{1},\mathbf{x}_{2},z_0,z_1,z_2,h_0,\lambda_2\Big) &=& 2i Q\, \textrm{K}_0\!\left(Q X\right)\Bigg\{z_1(1\!-\!z_1) \left[ 1\!-\!\frac{z_2}{(1\!-\!z_1)}\left(\frac{1\!-\!2 h_0\, \lambda_2}{2}\right)\right]\,
\frac{\varepsilon_{\lambda_2}^{*}\!\cdot\!\mathbf{x}_{20}}{{x}_{20}^2}\nonumber\\
& & -z_0(1\!-\!z_0) \left[ 1\!-\!\frac{z_2}{(1\!-\!z_0)}\left(\frac{1\!+\!2 h_0\, \lambda_2}{2}\right)\right]\,
\frac{\varepsilon_{\lambda_2}^{*}\!\cdot\!\mathbf{x}_{21}}{{x}_{21}^2}\Bigg\} \, . \label{NLO_L_wavefunction}
\end{eqnarray}
In order to get more compact expressions, the variable $X$ has been introduced, which is defined as
\begin{eqnarray}
X^2&=& z_1(1\!-\!z_1) \left(\mathbf{x}_{10}\!-\!\frac{z_2}{1\!-\!z_1} \mathbf{x}_{20}\right)^2+\frac{z_2(1\!-\!z_1\!-\!z_2)}{(1\!-\!z_1)}\, {x}_{20}^2\label{def_X_a}\\
&=& z_0(1\!-\!z_0) \left(\mathbf{x}_{01}\!-\!\frac{z_2}{1\!-\!z_0} \mathbf{x}_{21}\right)^2+\frac{z_2(1\!-\!z_0\!-\!z_2)}{(1\!-\!z_0)}\, {x}_{21}^2\label{def_X_b}\\
&=&z_1\, z_0\, {x}_{10}^2+z_2\, z_0\, {x}_{20}^2+ z_2\, z_1\, {x}_{21}^2\, .\label{def_X_3}
\end{eqnarray}
Those three expressions are indeed equal when $z_0\!+\!z_1\!+\!z_2=1$.

\subsection{Some comments about kinematical effects in the results\label{sec:comments}}

Let us discuss the variable $X$ and its various expressions in more details. The expression \eqref{def_X_a} is the one naturally obtained for the argument of the Bessel function $\textrm{K}_1(x)$ (or $\textrm{K}_0(x)$ in the longitudinal case) when calculating the contribution of the diagrams \ref{Fig:DiagNLOT}(a) and \ref{Fig:DiagNLOT}(c) thanks to the integrals \eqref{K1_NLO_2steps} and \eqref{K1_NLO_inst} (or \eqref{K0_NLO_2steps}), whereas
the expression \eqref{def_X_b} is the one naturally obtained from the diagrams \ref{Fig:DiagNLOT}(b) and \ref{Fig:DiagNLOT}(d). It is quite remarkable that those two complicated expressions are actually equal when $z_0\!+\!z_1\!+\!z_2=1$.

\subsubsection{Recoil effects}

If one makes the approximation $\mathbf{k}_1\simeq -\mathbf{k}_0$ in the energy denominator
$(Q^2+\mathbf{k}_0^2/z_0+\mathbf{k}_1^2/z_1+\mathbf{k}_2^2/z_2)$
of the momentum space expression \eqref{qqbarglue_T_wavefunction_mom}, one gets the simplification
\begin{equation}
\mathbf{x}_{10}\!-\!\frac{z_2}{1\!-\!z_1} \mathbf{x}_{20}\rightarrow  \mathbf{x}_{10}
\end{equation}
in the expression of $X$ in the contribution of the diagram \ref{Fig:DiagNLOT}(a) to the mixed space result. In momentum space, that approximation clearly corresponds to neglect some recoil effects and thus keep the same transverse momentum for the quark before and after the emission of the gluon. Hence, for the the diagram \ref{Fig:DiagNLOT}(a), it is natural to introduce the transverse point $\mathbf{x}_{0'}$, defined by
\begin{equation}
\mathbf{x}_{10'}=\mathbf{x}_{10}\!-\!\frac{z_2}{1\!-\!z_1} \mathbf{x}_{20}\, ,\label{would_be_position_of_quark}
\end{equation}
and interpret it as the would be position of the quark in the absence of gluon emission. Since the point $\mathbf{x}_{1}$ drops from that definition and $1\!-\!z_1=z_0\!+\!z_2$,
the equation \eqref{would_be_position_of_quark} is equivalent to
\begin{equation}
(z_0\!+\!z_2)\, \mathbf{x}_{0'}=z_0\, \mathbf{x}_{0}+ z_2\, \mathbf{x}_{2}\, .\label{conserv_positions}
\end{equation}
Notice that the points $\mathbf{x}_{i}$ (or $\mathbf{x}_{0'}$) are the positions (or would be position) of the partons at the time $x^+=0$ when they interact with the target, but not at other values of $x^+$.

On the other side, in light cone coordinates, the Poincar\'e algebra has a transverse Galilean subalgebra where  $\hat{{\cal P}}^-$ play the role of energy operator and thus $x^+$ of time, $\hat{{\cal P}}^+$ plays the role of mass and $\hat{{\cal P}}_{\perp}$ of momentum. This formalism then associates a transverse Galilean velocity $\mathbf{v}=\mathbf{k}/k^+$ to a free massless particle of momentum $k^+$ and $\mathbf{k}$. Then, the conservation of transverse momentum at the gluon emission vertex in the diagram \ref{Fig:DiagNLOT}(a) implies the relation
\begin{equation}
(z_0\!+\!z_2)\, \mathbf{v}_{0'}=z_0\, \mathbf{v}_{0}+ z_2\, \mathbf{v}_{2}\, ,\label{conserv_Gal_mom}
\end{equation}
between the transverse Galilean velocity of the parent quark $\mathbf{v}_{0'}$ and the ones of the daughter quark and gluon $\mathbf{v}_{0}$ and $\mathbf{v}_{2}$,  which is strongly reminiscent of the equation \eqref{conserv_positions}.

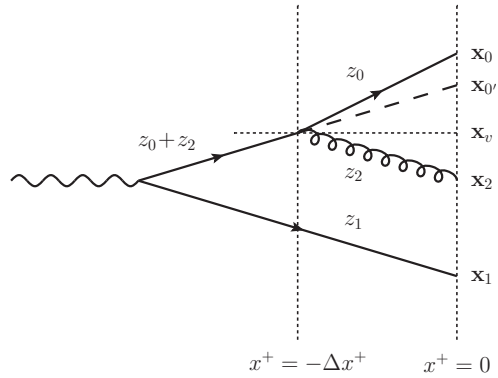
\begin{figure}
\setbox1\hbox to 10cm{
\fcolorbox{white}{white}{
  \begin{picture}(402,242) (9,-9)
    \SetWidth{1.0}
    \SetColor{White}
    \EBox(10,-8)(410,232)
    \SetWidth{1.4}
    \SetColor{Black}
    \Photon(10,122)(90,122){3.5}{4}
    \Line[arrow,arrowpos=0.5,arrowlength=5.667,arrowwidth=2.267,arrowinset=0.2](90,122)(190,152)
    \Line[arrow,arrowpos=0.5,arrowlength=5.667,arrowwidth=2.267,arrowinset=0.2](190,152)(290,202)
    \Gluon(190,152)(290,122){4}{8}
    \Line[arrow,arrowpos=0.5,arrowlength=5.667,arrowwidth=2.267,arrowinset=0.2](90,122)(290,62)
    \Line[dash,dashsize=10](190,152)(290,182)
    \SetWidth{1.0}
    \Line[dash,dashsize=2](290,232)(290,22)
    \Line[dash,dashsize=2](190,232)(190,22)
    \Line[dash,dashsize=2](150,152)(290,152)
    \Text(270,2)[lb]{\Large{\Black{$x^+=0$}}}
    \Text(160,2)[lb]{\Large{\Black{$x^+=-\Delta x^+$}}}
    \Text(300,147)[lb]{\Large{\Black{$\mathbf{x}_{v}$}}}
    \Text(300,177)[lb]{\Large{\Black{$\mathbf{x}_{0'}$}}}
    \Text(300,197)[lb]{\Large{\Black{$\mathbf{x}_{0}$}}}
    \Text(300,117)[lb]{\Large{\Black{$\mathbf{x}_{2}$}}}
    \Text(300,57)[lb]{\Large{\Black{$\mathbf{x}_{1}$}}}
    \Text(221,186)[lb]{\Large{\Black{$z_{0}$}}}
    \Text(220,122)[lb]{\Large{\Black{$z_{2}$}}}
    \Text(220,92)[lb]{\Large{\Black{$z_{1}$}}}
    \Text(90,142)[lb]{\Large{\Black{$z_{0}\!+\!z_2$}}}
  \end{picture}
}
}
\begin{center}
\hspace{-3cm}\resizebox*{6cm}{!}{\box1}
\caption{\label{Fig:Recoil}Geometric picture for the recoil effects found for the diagram \ref{Fig:DiagNLOT}(a) in mixed space.}
\end{center}
\end{figure}
In our perturbative description of the projectile, the partons should be free streaming in the transverse plane between two successive interaction vertices, or between the last vertex and the interaction with the target at $x^+=0$. Hence, geometrically, the diagram  \ref{Fig:DiagNLOT}(a) looks as in Fig.\ref{Fig:Recoil}. In the classical limit, there is a relation
\begin{equation}
\mathbf{x}_{i}=\mathbf{x}_{v}+\big(\Delta x^+\big)\,  \mathbf{v}_{i}\, ,\label{free_stream}
\end{equation}
for $i= 0$, $2$ or $0'$, where the emission of the gluon by the quark happens at a time $x^+=-\Delta x^+$ and a transverse position $\mathbf{x}_{v}$. That relation \eqref{free_stream} is sufficient to make the equations \eqref{conserv_positions} and \eqref{conserv_Gal_mom} equivalent. Beyond the classical limit, a relation of the type \eqref{free_stream} should hold when using a basis of wave packets to diagonalize the free hamiltonian instead of the momentum space or of the mixed space Fock states. In that case, the formula \eqref{free_stream} would relate the position of the center of the wave packet to the average momentum.

Hence, both by tracking the origin of the expression \eqref{would_be_position_of_quark} in our calculations and by discussing the Galilean subalgebra of the Poincar\'e algebra on the light front, one finds that
 the relation \eqref{conserv_positions} is the position space version of recoil effects associated to the gluon emission. This discussion of course generalizes to other diagrams and to the expression \eqref{def_X_b}.

Generically, for any initial state parton cascade described in mixed space, we thus know how to take recoil effects into account at each splitting. Consider a parton which splits into two in this cascade, which has a longitudinal momentum fraction $z_{\textrm{parent}}$ and which would be at the position $\mathbf{x}_{\textrm{parent}}$ at $x^+=0$ if it was not splitting before. Then, the two daughter partons must have longitudinal momentum fractions $z_{\textrm{d}1}$ and $z_{\textrm{d}2}$ and positions $\mathbf{x}_{\textrm{d}1}$ and $\mathbf{x}_{\textrm{d}2}$ at $x^+=0$ satisfying
\begin{eqnarray}
z_{\textrm{parent}}&=& z_{\textrm{d}1}+ z_{\textrm{d}2}\label{long_recoil}\\
z_{\textrm{parent}}\: \mathbf{x}_{\textrm{parent}}&=& z_{\textrm{d}1}\: \mathbf{x}_{\textrm{d}1}+ z_{\textrm{d}2}\: \mathbf{x}_{\textrm{d}2}\label{trans_recoil}\, ,
\end{eqnarray}
in order to conserve longitudinal and transverse momentum respectively.

\subsubsection{Formation time of multi-parton fluctuations}

Thanks to that discussion of recoil effects, it is natural to rewrite the equations \eqref{def_X_a} and \eqref{def_X_b} as
\begin{eqnarray}
X^2&=& z_1(1\!-\!z_1)\, {x}_{10'}^2+\frac{z_2\, z_0}{(z_2\!+\!z_0)}\, {x}_{20}^2\label{def_X_bis_a}\\
&=& z_0(1\!-\!z_0) \, {x}_{1'0}^2+\frac{z_2\, z_1}{(z_2\!+\!z_1)}\, {x}_{21}^2\, ,\label{def_X_bis_b}
\end{eqnarray}
where $\mathbf{x}_{1'}=(z_1\, \mathbf{x}_{1}\!+\!z_2\, \mathbf{x}_{2})/(z_1\!+\!z_2)$ is the would be position of the antiquark at $x^+=0$ in the absence of gluon emission in the diagram \ref{Fig:DiagNLOT}(b). In the expression \eqref{def_X_bis_a}, which is obtained for the diagram \ref{Fig:DiagNLOT}(a), the first term is obviously associated with the photon to dipole splitting and is a direct analog to the term appearing in the argument of the Bessel functions in the LO wave functions \eqref{LO_T_wavefunction} and \eqref{LO_L_wavefunction}, whereas the second term is associated with the emission of the gluon by the quark. And similarly, in the expression \eqref{def_X_bis_b} obtained from the diagram \ref{Fig:DiagNLOT}(b), the first term is again associated with the photon to dipole splitting and the second to the gluon emission by the antiquark.

It has often proven useful for the physical interpretation of perturbative results to associate a formation time (see \emph{e.g.} \cite{Dokshitzer:1991wu}) to a parton splitting. The formation time is basically the time it takes before the two daughter particles are away enough from each other compared to their wavelength so that they lose their quantum coherence and can thus act independently from each other. For the splitting of a parton of momentum $(k^+_{d1}\!+\!k^+_{d2},\mathbf{0})$ into two daughter partons of momentum $(k^+_{d1},\mathbf{k})$ and $(k^+_{d2},-\mathbf{k})$, the  formation time writes
\begin{equation}
\tau_{\textrm{form}}= \frac{2\, k^+_{d1}\, k^+_{d2}}{(k^+_{d1}\!+\!k^+_{d2})\, \mathbf{k}^2}\label{usual_form_time}\, .
\end{equation}

Up to a factor $2 q^+$, each of the terms in the expressions \eqref{def_X_bis_a} and \eqref{def_X_bis_b} is then the mixed space version of the formation time associated to a parton splitting, with the inverse of the relative transverse momentum of the daughters replaced by the their transverse distance at $x^+=0$.
The quantity $2 q^+\, X^2$ can thus be interpreted as the sum of the formation times associated with the two vertices of the diagram \ref{Fig:DiagNLOT}(a) (resp.  \ref{Fig:DiagNLOT}(b)) if one uses the expression \eqref{def_X_bis_a} (resp. \eqref{def_X_bis_b}).

A further hint for this interpretation is the following. The wave functions \eqref{NLO_T_wavefunction} and \eqref{NLO_L_wavefunction} are exponentially suppressed at large $Q X$ due to the Bessel functions, so that effectively, one has the restriction
\begin{equation}
2 q^+\, X^2 \lesssim \frac{2 q^+}{Q^2} \equiv \tau_{\gamma^*}\label{form_time_vs_lifetime}\, ,
\end{equation}
where $\tau_{\gamma^*}$ is the lifetime of the virtual photon. This inequality means that the two successive splittings in the diagrams \ref{Fig:DiagNLOT}(a) and \ref{Fig:DiagNLOT}(b) should happen within the lifetime of the parent virtual photon in order to give a relevant contribution. In the LO  wave functions \eqref{LO_T_wavefunction} and \eqref{LO_L_wavefunction}, the same interpretation holds. The only difference is the presence of a single splitting vertex instead of two.

At this point, we have understood the physics behind the expressions \eqref{def_X_a} and \eqref{def_X_b}, related to the diagrams \ref{Fig:DiagNLOT}(a) and \ref{Fig:DiagNLOT}(b) respectively, and also the argument of the Bessel functions in the LO wave functions. The first surprise is then that the expressions \eqref{def_X_a} and \eqref{def_X_b} are actually equal (provided recoil effects are not neglected), both reducing to the simple formula \eqref{def_X_3}. The second surprise is that even the instantaneous diagrams \ref{Fig:DiagNLOT}(c) and \ref{Fig:DiagNLOT}(d) give Bessel functions with the same argument, despite the fact that the interpretation of $2 q^+\, X^2$ as the sum of the formation times for two successive splittings does not make sense in those cases. One is thus led to the conclusion that $2 q^+\, X^2$ is the global formation time for a quark-antiquark-gluon system from a photon, and that it does not depend on the production mechanism but just on the final state of the cascade at $x^+=0$. Notice that none of the three terms in the expression \eqref{def_X_3}, taken individually, has an interpretation as a formation time for one splitting: the recoil effects are lacking in the first term, and the longitudinal momentum fraction of the parent is missing in the denominator of the second and third terms.
The final expression \eqref{def_X_3} for $X^2$ is fully symmetric with respect to permutations of the particles. The formation time is indeed a purely kinematical quantity, independent of the nature of the particles and of dynamical properties of the cascade like the color flow.

When analysing the structure of the perturbative contributions to the quark, antiquark plus $n$ gluons component of the photon wave function, one can see that our previous discussion generalizes. At any order in light-front perturbation theory, there should be a $\textrm{K}_1(Q\, X_n)$ (resp. $\textrm{K}_0(Q\, X_n)$) Bessel function which factorizes
from the $n$-partons component of the transverse (resp. longitudinal) photon wave function, with a common variable $X_n$ for all diagrams leading to the same $n$-partons Fock state.
And the mixed space expression for the formation time of a system of $n$ partons should write
\begin{equation}
\tau_{\textrm{form, }n\textrm{ part}}=2 q^+\, X_n ^2 =2 q^+ \sum^{n\!-\!1}_{\begin{array}{c}
                                                      {\scriptstyle i,j=0} \\
                                                      {\scriptstyle i< j}
                                                    \end{array}}
z_i\, z_j\, {x}_{ij}^2\, .\label{form_time_conjecture}
\end{equation}
For the $2$ and $3$ partons cases, that formula \eqref{form_time_conjecture} is indeed  given by our perturbative calculations. In the appendix \ref{sec:4part_form_time}, hints for the validity of the expression \eqref{form_time_conjecture} for the $4$ particles case are provided.

\subsection{Virtual corrections\label{sec:virt_corr}}

So far, we have calculated the quark-antiquark-gluon component but not the quark-antiquark component of the NLO virtual photon wave-functions \eqref{full_NLO_wavefunction}. The direct calculation of the latter in light-front perturbation theory would involve loop diagrams and the renormalization of the theory, which is particularly cumbersome in this context. However, those NLO quark-antiquark components can be easily determined according to a prescription based on probability conservation (see \emph{e.g.} \cite{Mueller:1993rr,Balitsky:1995ub}).
At LO, the normalization of the wave-function \eqref{LO_wavefunction} is non-trivial because we have dropped its colorless components, containing for example the photon itself or a lepton pair. Then,  at NLO, QCD corrections are allowed, but not electroweak corrections. Hence, the global normalization of the colorfull sector of the photon wave-functions is the same at LO and at NLO with respect to QCD, so that
\begin{eqnarray}
& &\!\!\!\!\!\!\!\!{\phantom{\Big\rangle}}_{LO}\Big\langle {\gamma_{T,L}^*\big(q'^+,Q^2,(\lambda)\big)}_H \Big|{\gamma_{T,L}^*\big(q^+,Q^2,(\lambda)\big)}_H\Big\rangle_{LO}
=\!\!\!\!{\phantom{\Big\rangle}}_{NLO}\Big\langle {\gamma_{T,L}^*\big(q'^+,Q^2,(\lambda)\big)}_H \Big|{\gamma_{T,L}^*\big(q^+,Q^2,(\lambda)\big)}_H\Big\rangle_{NLO}\nonumber\\
& &\qquad  =\!\!\!\! {\phantom{\Big\rangle}}_{q\bar{q}}\Big\langle {\gamma_{T,L}^*\big(q'^+,Q^2,(\lambda)\big)}_H \Big|{\gamma_{T,L}^*\big(q^+,Q^2,(\lambda)\big)}_H\Big\rangle_{q\bar{q}}+\!\!\!\!{\phantom{\Big\rangle}}_{q\bar{q}g}\Big\langle {\gamma_{T,L}^*\big(q'^+,Q^2,(\lambda)\big)}_H \Big|{\gamma_{T,L}^*\big(q^+,Q^2,(\lambda)\big)}_H\Big\rangle_{q\bar{q}g}
\, .\label{Normalization_conservation}
\end{eqnarray}

From the relation \eqref{LO_wavefunction}, one gets the normalization of the wave-functions at LO
\begin{eqnarray}
& &\!\!\!\!\!\!\!\!{\phantom{\Big\rangle}}_{LO}\Big\langle {\gamma_{T,L}^*\big(q'^+,Q^2,(\lambda)\big)}_H \Big|{\gamma_{T,L}^*\big(q^+,Q^2,(\lambda)\big)}_H\Big\rangle_{LO}\nonumber\\
& &\qquad =2\pi\, \delta\left(\frac{q'^+}{q^+}\!-\!1\right)  \frac{2N_c\, \alpha_{em}}{(2\pi)^2}\sum_f e_f^2   \int \textrm{d}^2\mathbf{x}_{0} \int \textrm{d}^2\mathbf{x}_{1}\, \int_0^1 \textrm{d} z_1\,  \sum_{h_0}  \: \left|\Phi^{LO}_{T,L}\Big(\mathbf{x}_{0},\mathbf{x}_{1},1\!-\!z_1,z_1,(h_0),(\lambda)\Big)\right|^2\, .\label{norm_LO}
\end{eqnarray}
And the normalization of the NLO quark-antiquark-gluon component is obtained from Eq.\eqref{qqbarglue_wavefunction} as
\begin{eqnarray}
& &\!\!\!\!\!\!\!\!{\phantom{\Big\rangle}}_{q\bar{q}g}\Big\langle {\gamma_{T,L}^*\big(q'^+,Q^2,(\lambda)\big)}_H \Big|{\gamma_{T,L}^*\big(q^+,Q^2,(\lambda)\big)}_H\Big\rangle_{q\bar{q}g}
  \qquad =2\pi\, \delta\left(\frac{q'^+}{q^+}\!-\!1\right)  \frac{2N_c\, \alpha_{em}}{(2\pi)^2}\sum_f e_f^2   \int \textrm{d}^2\mathbf{x}_{0} \int \textrm{d}^2\mathbf{x}_{1}\, \int_0^1 \textrm{d} z_1\,\nonumber\\
& &\qquad\qquad\qquad\times \left(1\!-\!\frac{1}{N_c^2}\right) \, \bar{\alpha} \int_0^{1\!-\!z_1} \frac{\textrm{d}z_2}{z_2}\,  \int \frac{\textrm{d}^2\mathbf{x}_{2}}{2\pi}\,
  \sum_{h_0,\lambda_2}  \: \left|\Phi^{q\bar{q}g}_{T,L}\Big(\mathbf{x}_{0},\mathbf{x}_{1},\mathbf{x}_{2},1\!-\!z_1\!-\!z_2,z_1,z_2,h_0,\lambda_2,(\lambda)\Big)\right|^2\, ,\label{norm_qqbarglue}
\end{eqnarray}
with the notation
\begin{equation}
\bar{\alpha} = \frac{N_c}{\pi}\, \alpha_s= \frac{N_c\, g^2}{(2\pi)^2}\label{alpha_bar}\, .
\end{equation}
The expression \eqref{norm_qqbarglue} has divergences for $\mathbf{x}_{2}\rightarrow \mathbf{x}_{0}$, for $\mathbf{x}_{2}\rightarrow \mathbf{x}_{1}$ and for $z_2\rightarrow 0$, which should be regularized in some way. However, we will keep the regularization implicit here, for convenience. The first two divergences will disappear in the expressions for the virtual photon cross section, and we will come back in section \ref{sec:subtract_LL} on the $z_2\rightarrow 0$ divergence, associated with the low $x$ factorization and evolution.

The quark-antiquark component of the NLO virtual photon wave-functions can be parameterized in a similar way as the LO one \eqref{LO_wavefunction}
\begin{eqnarray}
\left|{\gamma_{T,L}^*\big(q^+,Q^2,(\lambda)\big)}_H\right\rangle_{q\bar{q}} &=&\frac{e}{2} \int_0^1 \frac{\textrm{d}z_0}{\sqrt{z_0}} \int_0^1 \frac{\textrm{d}z_1}{\sqrt{z_1}}\, \delta(z_0\!+\!z_1\!-\!1)  \int \frac{\textrm{d}^2\mathbf{x}_{0}}{(2\pi)^2}
 \int \frac{\textrm{d}^2\mathbf{x}_{1}}{(2\pi)^2}  \sum_{h_0}  \: \Phi^{q\bar{q}}_{T,L}\Big(\mathbf{x}_{0},\mathbf{x}_{1},z_0,z_1,h_0,(\lambda)\Big)
\nonumber\\
 & &\quad \times \sum_f e_f \sum_{A_0} \: b^\dag(\mathbf{x}_0,k^+_0,h_0,A_0,f)  d^\dag(\mathbf{x}_1,k^+_1,-h_0,A_0,f)  |0\rangle\, .\label{qqbar_wavefunction}
\end{eqnarray}
This leads to an expression analog to \eqref{norm_LO} for the normalization of that component.

Then, let us follow the standard prescription \cite{Mueller:1993rr,Balitsky:1995ub} and assume that the probability conservation relation \eqref{Normalization_conservation} is actually valid at the level of the integrand, \emph{i.e.} for each value of $\mathbf{x}_{0}$, $\mathbf{x}_{1}$ and $z_1$
\begin{eqnarray}
& &\!\!\!\!\!\!\!\!\sum_{h_0}  \: \left|\Phi^{LO}_{T,L}\Big(\mathbf{x}_{0},\mathbf{x}_{1},1\!-\!z_1,z_1,(h_0),(\lambda)\Big)\right|^2
=\sum_{h_0}  \: \left|\Phi^{q\bar{q}}_{T,L}\Big(\mathbf{x}_{0},\mathbf{x}_{1},1\!-\!z_1,z_1,h_0,(\lambda)\Big)\right|^2\nonumber\\
& &\qquad\qquad\qquad +\left(1\!-\!\frac{1}{N_c^2}\right) \, \bar{\alpha} \int_0^{1\!-\!z_1} \frac{\textrm{d}z_2}{z_2}\,  \int \frac{\textrm{d}^2\mathbf{x}_{2}}{2\pi}\,
  \sum_{h_0,\lambda_2}  \: \left|\Phi^{q\bar{q}g}_{T,L}\Big(\mathbf{x}_{0},\mathbf{x}_{1},\mathbf{x}_{2},1\!-\!z_1\!-\!z_2,z_1,z_2,h_0,\lambda_2,(\lambda)\Big)\right|^2
\, .\label{Normalization_conservation_unintegrated}
\end{eqnarray}


\section{Photon cross sections and impact factors at NLO\label{sec:crossSection}}

\subsection{From wave-functions to cross sections}

Thanks to the general formula \eqref{OptTh}, one writes the longitudinal photon cross section $\sigma_{L}^{\gamma}\left[{\cal A}\right]$ on a given classical shockwave field ${\cal A}^-_a$ at NLO accuracy as
\begin{eqnarray}
\sigma_{L}^{\gamma}\left[{\cal A}\right]&=&  \frac{1}{(2\pi)\, \delta\left(\frac{{q'}^+}{q^+}\!-\!1\right)}\:  \textrm{Re}\left({\phantom{\Big\rangle}}_{NLO}\Big\langle {\gamma_{L}^*\big(q'^+,Q^2\big)}_H \Big| 1\!-\! \hat{S}_E \Big|{\gamma_{L}^*\big(q^+,Q^2\big)}_H\Big\rangle_{NLO}\;\;\right)\nonumber\\
&=& \frac{1}{(2\pi)\, \delta\left(\frac{{q'}^+}{q^+}\!-\!1\right)}\:  \Bigg\{{\phantom{\Big\rangle}}_{LO}\Big\langle {\gamma_{L}^*\big(q'^+,Q^2\big)}_H \Big|{\gamma_{L}^*\big(q^+,Q^2\big)}_H\Big\rangle_{LO}\nonumber\\
& & - \!\!\!\!{\phantom{\Big\rangle}}_{q\bar{q}}\Big\langle {\gamma_{L}^*\big(q'^+,Q^2\big)}_H \Big|\hat{S}_E \Big|{\gamma_{L}^*\big(q^+,Q^2\big)}_H\Big\rangle_{q\bar{q}}\;\; - \!\!\!\!{\phantom{\Big\rangle}}_{q\bar{q}g}\Big\langle {\gamma_{L}^*\big(q'^+,Q^2\big)}_H \Big|\hat{S}_E \Big|{\gamma_{L}^*\big(q^+,Q^2\big)}_H\Big\rangle_{q\bar{q}g}
\Bigg\}
\, .\label{sigma_L_1}
\end{eqnarray}
Here the relation \eqref{Normalization_conservation} has been used. The eikonal scattering operator $\hat{S}_E$ does not change the partonic content of the projectile and has purely real matrix elements for colorless projectiles. That is why one can drop the real part operator $\textrm{Re}$ and split the $\hat{S}_E$ matrix element at NLO into the $q\bar{q}$ and $q\bar{q}g$ sector contributions. The transverse virtual photon cross section $\sigma_{T}^{\gamma}$ is defined to be unpolarized. Hence, the expression for $\sigma_{T}^{\gamma}\left[{\cal A}\right]$ at NLO accuracy is analog to the expression \eqref{sigma_L_1} for $\sigma_{L}^{\gamma}\left[{\cal A}\right]$, except that one has to average over the photon helicity $\lambda$.

Thanks to the relations \eqref{action_of_S_E} and \eqref{qqbar_wavefunction}  the NLO matrix elements of $\hat{S}_E$ writes in the $q\bar{q}$ sector
\begin{eqnarray}
& &\!\!\!\!\!\!\!\!{\phantom{\Big\rangle}}_{q\bar{q}}\Big\langle {\gamma_{T,L}^*\big(q'^+,Q^2,(\lambda)\big)}_H \Big|\hat{S}_E\Big|{\gamma_{T,L}^*\big(q^+,Q^2,(\lambda)\big)}_H\Big\rangle_{q\bar{q}}\nonumber\\
& &\quad =2\pi\, \delta\left(\frac{q'^+}{q^+}\!-\!1\right)  \frac{2N_c\, \alpha_{em}}{(2\pi)^2}\sum_f e_f^2   \int \textrm{d}^2\mathbf{x}_{0} \int \textrm{d}^2\mathbf{x}_{1} \; {\cal S}_{01}\!\left[{\cal A}\right] \int_0^1 \textrm{d} z_1\:
  \sum_{h_0}  \: \left|\Phi^{q\bar{q}}_{T,L}\Big(\mathbf{x}_{0},\mathbf{x}_{1},1\!-\!z_1,z_1,h_0,(\lambda)\Big)\right|^2\, ,\label{SE_Matrix_elem_qqbar_1}
\end{eqnarray}
with the notation
\begin{equation}
{\cal S}_{ij}\!\left[{\cal A}\right] = \frac{1}{N_c}\: \textrm{tr}\left(U\!\left[{\cal A}\right]\!(\mathbf{x}_0)\; U^\dag\!\left[{\cal A}\right]\!(\mathbf{x}_1)\right) \label{dipole_S_matrix}
\end{equation}
for the so-called dipole S-matrix.
Because of the relation \eqref{Normalization_conservation_unintegrated}, the equation \eqref{SE_Matrix_elem_qqbar_1} can be rewritten as
\begin{eqnarray}
& &\!\!\!\!\!\!\!\!{\phantom{\Big\rangle}}_{q\bar{q}}\Big\langle {\gamma_{T,L}^*\big(q'^+,Q^2,(\lambda)\big)}_H \Big|\hat{S}_E\Big|{\gamma_{T,L}^*\big(q^+,Q^2,(\lambda)\big)}_H\Big\rangle_{q\bar{q}}\nonumber\\
& &\quad =2\pi\, \delta\left(\frac{q'^+}{q^+}\!-\!1\right)  \frac{2N_c\, \alpha_{em}}{(2\pi)^2}\sum_f e_f^2   \int \textrm{d}^2\mathbf{x}_{0} \int \textrm{d}^2\mathbf{x}_{1} \; {\cal S}_{01}\!\left[{\cal A}\right] \int_0^1 \textrm{d} z_1\:
  \sum_{h_0}  \:\Bigg\{\left|\Phi^{LO}_{T,L}\Big(\mathbf{x}_{0},\mathbf{x}_{1},1\!-\!z_1,z_1,h_0,(\lambda)\Big)\right|^2\nonumber\\
& &\quad \quad\quad -  \left(1\!-\!\frac{1}{N_c^2}\right) \, \bar{\alpha} \int_0^{1\!-\!z_1} \frac{\textrm{d}z_2}{z_2}\,  \int \frac{\textrm{d}^2\mathbf{x}_{2}}{2\pi}\,
  \sum_{\lambda_2}  \: \left|\Phi^{q\bar{q}g}_{T,L}\Big(\mathbf{x}_{0},\mathbf{x}_{1},\mathbf{x}_{2},1\!-\!z_1\!-\!z_2,z_1,z_2,h_0,\lambda_2,(\lambda)\Big)\right|^2\Bigg\}
  \, .\label{SE_Matrix_elem_qqbar}
\end{eqnarray}

Using the relations \eqref{action_of_S_E} and \eqref{qqbarglue_wavefunction}, the calculation of the NLO matrix elements of $\hat{S}_E$ in the $q\bar{q}g$ sector gives
\begin{eqnarray}
& &\!\!\!\!\!\!{\phantom{\Big\rangle}}_{q\bar{q}g}\Big\langle {\gamma_{T,L}^*\big(q'^+,Q^2,(\lambda)\big)}_H \Big|\hat{S}_E\Big|{\gamma_{T,L}^*\big(q^+,Q^2,(\lambda)\big)}_H\Big\rangle_{q\bar{q}g} =2\pi\, \delta\left(\frac{q'^+}{q^+}\!-\!1\right)  \frac{4\, \alpha_{em}\, \bar{\alpha}}{N_c\, (2\pi)^2}\sum_f e_f^2   \int \textrm{d}^2\mathbf{x}_{0} \int \textrm{d}^2\mathbf{x}_{1} \nonumber\\
& & \qquad \qquad\qquad \times \int \frac{\textrm{d}^2\mathbf{x}_{2}}{2\pi} \int_0^1 \textrm{d} z_1 \int_0^{1\!-\!z_1} \frac{\textrm{d}z_2}{z_2}\,
    \: \Big\{\sum_{a,b}  \left[V\left[{\cal A}\right](\mathbf{x}_2)\right]_{b a}
\: \textrm{tr}\left(U\!\left[{\cal A}\right]\!(\mathbf{x}_0)\; T^a \;U^\dag\!\left[{\cal A}\right]\!(\mathbf{x}_1)\: T^b\right)\Big\}\nonumber\\
& & \qquad \qquad\qquad \times
  \sum_{h_0,\lambda_2}  \: \left|\Phi^{q\bar{q}g}_{T,L}\Big(\mathbf{x}_{0},\mathbf{x}_{1},\mathbf{x}_{2},1\!-\!z_1\!-\!z_2,z_1,z_2,h_0,\lambda_2,(\lambda)\Big)\right|^2\nonumber\\
& &\qquad\qquad= 2\pi\, \delta\left(\frac{q'^+}{q^+}\!-\!1\right)  \frac{2N_c\, \alpha_{em}}{(2\pi)^2}\sum_f e_f^2   \int \textrm{d}^2\mathbf{x}_{0} \int \textrm{d}^2\mathbf{x}_{1} \int_0^1 \textrm{d} z_1\, \bar{\alpha}  \int \frac{\textrm{d}^2\mathbf{x}_{2}}{2\pi} \int_0^{1\!-\!z_1}\frac{\textrm{d}z_2}{z_2}\nonumber\\
& & \qquad \qquad\qquad \times \Bigg[{\cal S}_{02}\!\left[{\cal A}\right]\; {\cal S}_{21}\!\left[{\cal A}\right]  \!-\!\frac{1}{N_c^2}\, {\cal S}_{01}\!\left[{\cal A}\right]\Bigg]
\, \sum_{h_0,\lambda_2}  \: \left|\Phi^{q\bar{q}g}_{T,L}\Big(\mathbf{x}_{0},\mathbf{x}_{1},\mathbf{x}_{2},1\!-\!z_1\!-\!z_2,z_1,z_2,h_0,\lambda_2,(\lambda)\Big)\right|^2
 \, ,\label{SE_Matrix_elem_qqbarglue}
\end{eqnarray}
where the second expression is obtained thanks to the identities
\begin{eqnarray}
 \left[V\left[{\cal A}\right](\mathbf{x})\right]_{b a}&=&2\;\textrm{tr}\left(U\!\left[{\cal A}\right]\!(\mathbf{x})\; T^a \;U^\dag\!\left[{\cal A}\right]\!(\mathbf{x})\: T^b\right)\\
 \textrm{and} \qquad \qquad \sum_a\:  \big(T^a\big)_{A B}\; \big(T^a\big)_{C D} &=&  \frac{1}{2}\, \delta_{A,D}\, \delta_{B,C} -  \frac{1}{2 N_c}\, \delta_{A,B}\, \delta_{C,D}\, .
\end{eqnarray}

Finally, inserting the results \eqref{norm_LO}, \eqref{SE_Matrix_elem_qqbar} and \eqref{SE_Matrix_elem_qqbarglue} into the equation \eqref{sigma_L_1} or its analog for the transverse photon cross section,  one gets the photon cross sections for a given classical gluon shockwave
\begin{eqnarray}
\sigma_{T,L}^{\gamma}\left[{\cal A}\right]&=& 2\; \frac{2N_c\, \alpha_{em}}{(2\pi)^2}\sum_f e_f^2   \int \textrm{d}^2\mathbf{x}_{0} \int \textrm{d}^2\mathbf{x}_{1} \int_0^1 \textrm{d} z_1\, \Bigg\{\Big[1- {\cal S}_{01}\left[{\cal A}\right] \Big]\; \mathcal{I}_{T,L}^{LO}(\mathbf{x}_{0},\mathbf{x}_{1},1\!-\!z_1,z_1)\nonumber\\
& & + \bar{\alpha}  \int \frac{\textrm{d}^2\mathbf{x}_{2}}{2\pi} \int_0^{1\!-\!z_1}\frac{\textrm{d}z_2}{z_2}
\Big[ {\cal S}_{01}\left[{\cal A}\right]- {\cal S}_{02}\left[{\cal A}\right]\, {\cal S}_{21}\left[{\cal A}\right]\Big]\;  \mathcal{I}_{T,L}^{NLO}(\mathbf{x}_{0},\mathbf{x}_{1},\mathbf{x}_{2},1\!-\!z_1\!-\!z_2,z_1,z_2)
\Bigg\}\, ,\label{sigma_TL_A}
\end{eqnarray}
with the impact factors defined as
\begin{eqnarray}
\mathcal{I}_{L}^{LO}(\mathbf{x}_{0},\mathbf{x}_{1},z_0,z_1)&=&\frac{1}{2}\: \sum_{h_0} \left|\Phi^{LO}_{L}\Big(\mathbf{x}_{0},\mathbf{x}_{1},z_0,z_1\Big)\right|^2\label{ImpFact_LO_L_def}\\
\mathcal{I}_{T}^{LO}(\mathbf{x}_{0},\mathbf{x}_{1},z_0,z_1)&=&\frac{1}{4}\: \sum_{h_0,\lambda} \left|\Phi^{LO}_{T}\Big(\mathbf{x}_{0},\mathbf{x}_{1},z_0,z_1,h_0,\lambda\Big)\right|^2\label{ImpFact_LO_T_def}\\
\mathcal{I}_{L}^{NLO}(\mathbf{x}_{0},\mathbf{x}_{1},\mathbf{x}_{2},z_0,z_1,z_2)&=&\frac{1}{2}\: \sum_{h_0, \lambda_2} \left|\Phi^{q\bar{q}g}_{L}\Big(\mathbf{x}_{0},\mathbf{x}_{1},\mathbf{x}_{2},z_0,z_1,z_2,h_0,\lambda_2\Big)\right|^2\label{ImpFact_NLO_L_def}\\
\mathcal{I}_{T}^{NLO}(\mathbf{x}_{0},\mathbf{x}_{1},\mathbf{x}_{2},z_0,z_1,z_2)&=&\frac{1}{4}\: \sum_{h_0,\lambda, \lambda_2} \left|\Phi^{q\bar{q}g}_{T}\Big(\mathbf{x}_{0},\mathbf{x}_{1},\mathbf{x}_{2},z_0,z_1,z_2,h_0,\lambda,\lambda_2\Big)\right|^2\label{ImpFact_NLO_T_def}
\, .
\end{eqnarray}

Then, from the expressions \eqref{LO_T_wavefunction} and \eqref{LO_L_wavefunction}, one recovers of course the classic results of Nikolaev and Zakharov \cite{Nikolaev:1990ja} for the LO impact factors for the dipole factorization
\begin{eqnarray}
\mathcal{I}_{L}^{LO}(\mathbf{x}_{0},\mathbf{x}_{1},z_0,z_1)&=& 4 Q^2 z_0^2 z_1^2 \, \textrm{K}_0^2\!\left(Q\sqrt{z_0 z_1 {x}_{01}^2}\right)\label{ImpFact_LO_L}\\
\mathcal{I}_{T}^{LO}(\mathbf{x}_{0},\mathbf{x}_{1},z_0,z_1)&=& \big[z_0^2+z_1^2\big] z_0 z_1 Q^2 \textrm{K}_1^2\!\left(Q\sqrt{z_0 z_1 {x}_{01}^2}\right)\label{ImpFact_LO_T}\, .
\end{eqnarray}
The NLO impact factors, which constitute the main results of the present work, are calculated using the intermediate results \eqref{NLO_T_wavefunction} and \eqref{NLO_L_wavefunction}, and write
\begin{eqnarray}
\mathcal{I}_{L}^{NLO}(\mathbf{x}_{0},\mathbf{x}_{1},\mathbf{x}_{2},z_0,z_1,z_2)&=&4 Q^2 \, \textrm{K}_0^2\!\left(QX\right) \Bigg\{z_1^2 (1\!-\!z_1)^2 \frac{\,{\cal P}\!\left(\frac{z_2}{1\!-\!z_1}\right)}{{x}_{20}^2}
+z_0^2 (1\!-\!z_0)^2 \frac{\,{\cal P}\!\left(\frac{z_2}{1\!-\!z_0}\right)}{{x}_{21}^2}\nonumber\\
& & \qquad \qquad -2 z_1 (1\!-\!z_1) z_0 (1\!-\!z_0) \left[1\!-\!\frac{z_2}{2(1\!-\!z_1)}\!-\!\frac{z_2}{2(1\!-\!z_0)}\right] \left(\frac{\mathbf{x}_{20}\cdot\mathbf{x}_{21}}{{x}_{20}^2\; {x}_{21}^2}\right)
\Bigg\}
\label{ImpFact_NLO_L}
\end{eqnarray}
and
\begin{eqnarray}
& &\mathcal{I}_{T}^{NLO}(\mathbf{x}_{0},\mathbf{x}_{1},\mathbf{x}_{2},z_0,z_1,z_2)= \Bigg[\frac{Q X\, \textrm{K}_1\!\left(Q X\right)}{X^2}\Bigg]^2
\Bigg\{z_1^2 (1\!-\!z_1)^2 \big[z_1^2+(1\!-\!z_1)^2\big]  \left(\mathbf{x}_{10}\!-\!\frac{z_2}{1\!-\!z_1} \mathbf{x}_{20}\right)^2  \frac{\,{\cal P}\!\left(\frac{z_2}{1\!-\!z_1}\right)}{{x}_{20}^2}\nonumber\\
& &+z_0^2 (1\!-\!z_0)^2 \big[z_0^2+(1\!-\!z_0)^2\big]  \left(\mathbf{x}_{01}\!-\!\frac{z_2}{1\!-\!z_0} \mathbf{x}_{21}\right)^2  \frac{\,{\cal P}\!\left(\frac{z_2}{1\!-\!z_0}\right)}{{x}_{21}^2}\nonumber\\
& &+2 z_1 (1\!-\!z_1) z_0 (1\!-\!z_0) \big[z_1(1\!-\!z_0)+z_0(1\!-\!z_1)\big] \left[1\!-\!\frac{z_2}{2(1\!-\!z_1)}\!-\!\frac{z_2}{2(1\!-\!z_0)}\right]
\left(\mathbf{x}_{10}\!-\!\frac{z_2}{1\!-\!z_1} \mathbf{x}_{20}\right)\!\!\cdot\!\! \left(\mathbf{x}_{01}\!-\!\frac{z_2}{1\!-\!z_0} \mathbf{x}_{21}\right)
\left(\frac{\mathbf{x}_{20}\cdot\mathbf{x}_{21}}{{x}_{20}^2\; {x}_{21}^2}\right)\nonumber\\
& & + \frac{z_0\, z_1\,z_2^2\, (z_0\!-\!z_1)^2}{(1\!-\!z_1) (1\!-\!z_0)}\; \frac{\big(\mathbf{x}_{20}\wedge\mathbf{x}_{21}\big)^2}{{x}_{20}^2\; {x}_{21}^2}
+ z_0\, z_1^2\, z_2 \bigg[\frac{z_0\, z_1}{(1\!-\!z_1)}+\frac{(1\!-\!z_1)^2}{(1\!-\!z_0)} \bigg] \left(\mathbf{x}_{10}\!-\!\frac{z_2}{1\!-\!z_1} \mathbf{x}_{20}\right)\!\!\cdot\!\! \left(\frac{\mathbf{x}_{20}}{{x}_{20}^2}\right)\nonumber\\
& &
+ z_0^2\, z_1\, z_2 \bigg[\frac{z_0\, z_1}{(1\!-\!z_0)}+\frac{(1\!-\!z_0)^2}{(1\!-\!z_1)} \bigg] \left(\mathbf{x}_{01}\!-\!\frac{z_2}{1\!-\!z_0} \mathbf{x}_{21}\right)\!\!\cdot\!\! \left(\frac{\mathbf{x}_{21}}{{x}_{21}^2}\right)
+ \frac{z_0^2\, z_1^2\, z_2^2}{2} \bigg[\frac{1}{(1\!-\!z_1)^2}+\frac{1}{(1\!-\!z_0)^2} \bigg]
\Bigg\}
\, ,\label{ImpFact_NLO_T}
\end{eqnarray}
where $\mathbf{x}\wedge \mathbf{y}\equiv \epsilon^{ij}\,  \mathbf{x}^i\, \mathbf{y}^j$ and
\begin{equation}
{\cal P}(z)=\frac{1}{2} \left[1+ \left(1\!-\!z\right)^2\right]\label{Pgq_splitting}\, .
\end{equation}

The two first terms in the NLO impact factors \eqref{ImpFact_NLO_L} and \eqref{ImpFact_NLO_T} come from the squares of the diagram \ref{Fig:DiagNLOT}(a) and \ref{Fig:DiagNLOT}(b) respectively (or their analog for longitudinal photon). Since ${\cal P}(z)/z$ is the quark to gluon DGLAP splitting function, up to some color factor, those two terms correctly reproduce the first step in the LO DGLAP evolution of the virtual photon in the appropriate limits ${x}_{20}^2\rightarrow 0$ and ${x}_{21}^2\rightarrow 0$, relevant in the so-called resolved photon kinematics. By contrast, the first contribution to the more familiar DGLAP evolution of the target in DIS is not clearly apparent in the results \eqref{sigma_TL_A}, \eqref{ImpFact_NLO_L} and \eqref{ImpFact_NLO_T}, although it should be somehow present there.
In the limits ${x}_{20}^2\rightarrow 0$ and ${x}_{21}^2\rightarrow 0$, the two first terms in the NLO impact factors \eqref{ImpFact_NLO_L} and \eqref{ImpFact_NLO_T} have the singular behavior responsible for the logarithmic divergences already mentioned in the section \ref{sec:virt_corr}. However, those divergences are regulated in the cross sections $\sigma_{T,L}^{\gamma}\left[{\cal A}\right]$ \eqref{sigma_TL_A} by the property of the color transparency
\begin{equation}
   1-{\cal S}_{ij}\!\left[{\cal A}\right]\; \propto  {x}_{ij}^2 \quad \textrm{for} \quad \mathbf{x}_{j}\rightarrow \mathbf{x}_{i}\, .
\end{equation}
Hence, the integration over $\mathbf{x}_{2}$ in the formula \eqref{sigma_TL_A} does not need any regularization anymore.\\

The third term in the NLO impact factors \eqref{ImpFact_NLO_L} and \eqref{ImpFact_NLO_T} comes from the interference between the diagrams \ref{Fig:DiagNLOT}(a) and \ref{Fig:DiagNLOT}(b). However, in the transverse case \eqref{ImpFact_NLO_T}, the fourth term is an another contribution arising from the same interference. In intermediate stages of its calculation, this additional interference contribution is proportional to $(\mathbf{x}_{10'}\wedge\mathbf{x}_{1'0}) (\mathbf{x}_{20}\wedge\mathbf{x}_{21})$, where, following the notations of the section \ref{sec:comments}, $\mathbf{x}_{10'}$ (resp. $\mathbf{x}_{1'0}$) is the parent dipole vector in the diagram \ref{Fig:DiagNLOT}(a) (resp. \ref{Fig:DiagNLOT}(b)) including recoil effects. Notice that this contribution is the product of two terms which are odd by exchange of $\mathbf{x}_{1}$ and $z_1$ with $\mathbf{x}_{0}$ and $z_0$, so that it is globally even, as are other contributions to the impact factors \eqref{ImpFact_LO_L}, \eqref{ImpFact_LO_T}, \eqref{ImpFact_NLO_L} and \eqref{ImpFact_NLO_T}. Such an odd-odd contribution requires two vertices with a non-trivial tensor structure. In particular, the effective vertex \eqref{gammaL2qqbarVertexEff} or the instantaneous vertex \eqref{inst_gamma2qqbarg_vertex} are essentially scalar so that they cannot provide an odd product. All this explains why the odd-odd contributions start to appear at NLO for the transverse photon case, and not before NNLO for the longitudinal case. In addition, the antisymmetry of each factor implies that only interference terms can produce a non-vanishing odd-odd contribution. Recoil effects are also crucial for the appearance of the odd-odd contribution in the result \eqref{ImpFact_NLO_T}, by ensuring that $\mathbf{x}_{10'}\wedge\mathbf{x}_{1'0}\neq \mathbf{x}_{10}\wedge\mathbf{x}_{10}=0$.\\

Finally, the last term in the expression \eqref{ImpFact_NLO_T} comes from the square of the instantaneous diagrams  \ref{Fig:DiagNLOT}(c) and \ref{Fig:DiagNLOT}(d) and the fifth and sixth terms from interferences between instantaneous and non-instantaneous diagrams, which explains why they have no analog in the longitudinal photon case \eqref{ImpFact_NLO_L}.


\subsection{Subtracting high energy leading logs\label{sec:subtract_LL}}

In order to get the final result for the photon-target cross sections from \eqref{sigma_TL_A}, one still has to take the CGC average over the classical gluon field of the target and to deal with the logarithmic divergence at $z_2\rightarrow 0$. For generic field ${\cal A}^-_a$ of the target, the dipole S-matrix ${\cal S}_{ij}\!\left[{\cal A}\right]$ suffers from the so called rapidity divergence. Those two divergence are actually related and can be simultaneously regulated by a single cut-off preventing double counting of gluons. The idea is roughly to consider, in some particular frame, the right moving gluons as part of the projectile and the left moving gluons as part of the target. In practice, various choices of factorization scheme are possible, but the most convenient in our case is to use a cut-off $k^+_f$ in $k^+$, following \emph{e.g.} \cite{Balitsky:2008zz}. The CGC average is then restricted to fields ${\cal A}^-_a$ which have no modes of $k^+$ larger than $k^+_f$, and in the formula for the photon-target cross section the lower bound for the integration over $z_2$ is taken to be $z_f=k^+_f/q^+$.

The lowest possible value $k^+_{\min}$ for the cut-off $k^+_f$ is set by the target. Indeed, from the point of view of the target,  a radiated gluon should be a fluctuation of lifetime $1/k^+$ shorter than the lifetime of the typical non-perturbative constituents of the target. Assuming that each of those constituents carry a sizable fraction of the momentum $P^-$ of the target\footnote{In the case of a target nucleus, $P^-$ is defined as the momentum per nucleon rather than the total one.} and are associated to a typical transverse scale $Q_0$, that bound writes
\begin{equation}
\frac{1}{k^+}<\frac{1}{k^+_{\min}}=\frac{2\, P^-}{Q_0^2}  \quad \textrm{so that} \quad z=\frac{k^+}{q^+}> z_{\min} =  \frac{Q_0^2}{2\, P^- q^+}=x\,  \frac{Q_0^2}{Q^2}\label{z_min}\, .
\end{equation}
For a hadron target $Q_0$ should be a non-perturbative QCD scale, and for a large nucleus target $Q_0$ can be identified with the initial saturation scale $Q_s(0)$.

From the formula \eqref{sigma_TL_A}, it is obvious that the cut-off $z_f$ should be smaller than $1\!-\!z_1$. Hence, $z_f$ satisfies
\begin{equation}
x\,  \frac{Q_0^2}{Q^2} <z_f \lesssim 1\!-\!z_1 \label{z_range}\, .
\end{equation}
It is convenient to introduce the variable
\begin{equation}
Y_f^+= \log\left(\frac{k^+_f}{k^+_{\min}}\right) = \log\left(\frac{z_f\, Q^2}{x\, Q_0^2}\right) \label{Yfplus_def}\, ,
\end{equation}
analogous to but distinct from a rapidity range, in order to specify the range left for the evolution of the target. That variable is bounded by
\begin{equation}
 0<Y_f^+ \lesssim \log\left(\frac{(1\!-\!z_1)\, Q^2}{x\, Q_0^2}\right) \label{Yfplus_range}\, .
\end{equation}
 The CGC average over the classical gluon field, associated to the cut-off $k^+_f$, is then noted $\langle \dots \rangle_{Y_f^+}$.

All in all, one can deduce from the equation \eqref{sigma_TL_A} the final result for the transverse and longitudinal virtual photon cross sections
\begin{eqnarray}
\sigma_{T,L}^{\gamma}&=& 2\; \frac{2N_c\, \alpha_{em}}{(2\pi)^2}\sum_f e_f^2   \int \textrm{d}^2\mathbf{x}_{0} \int \textrm{d}^2\mathbf{x}_{1} \int_0^1 \textrm{d} z_1\, \Bigg\{ \mathcal{I}_{T,L}^{LO}(\mathbf{x}_{0},\mathbf{x}_{1},1\!-\!z_1,z_1)\nonumber\\
& & \times \Bigg[1- \left\langle {\cal S}_{01} \right\rangle_{Y_f^+} + \bar{\alpha} \log\left(\frac{1\!-\!z_1}{z_f}\right)   \int \frac{\textrm{d}^2\mathbf{x}_{2}}{2\pi} \frac{x_{01}^2}{x_{02}^2\, x_{21}^2}\: \left\langle {\cal S}_{01}\!-\! {\cal S}_{02} {\cal S}_{21} \right\rangle_{Y_f^+}
\Bigg]\;\nonumber\\
& & + \bar{\alpha}  \int \frac{\textrm{d}^2\mathbf{x}_{2}}{2\pi}\; \left\langle {\cal S}_{01}- {\cal S}_{02}\,{\cal S}_{21}\right\rangle_{Y_f^+} \int_{z_f}^{1\!-\!z_1}\frac{\textrm{d}z_2}{z_2}\;
  \Delta\mathcal{I}_{T,L}^{NLO}(\mathbf{x}_{0},\mathbf{x}_{1},\mathbf{x}_{2},z_1,z_2)
\Bigg\}\, ,\label{sigma_TL_Y_fplus}
\end{eqnarray}
where the high energy leading logs have been separated from the rest of the NLO corrections by defining
\begin{eqnarray}
\Delta\mathcal{I}_{T,L}^{NLO}(\mathbf{x}_{0},\mathbf{x}_{1},\mathbf{x}_{2},z_1,z_2)&=&
\mathcal{I}_{T,L}^{NLO}(\mathbf{x}_{0},\mathbf{x}_{1},\mathbf{x}_{2},1\!-\!z_1\!-\!z_2,z_1,z_2)
-\mathcal{I}_{T,L}^{NLO}(\mathbf{x}_{0},\mathbf{x}_{1},\mathbf{x}_{2},1\!-\!z_1,z_1,0)
\label{ImpFact_NLO_soft_regul}
\end{eqnarray}
and noticing that
\begin{eqnarray}
\mathcal{I}_{T,L}^{NLO}(\mathbf{x}_{0},\mathbf{x}_{1},\mathbf{x}_{2},1\!-\!z_1,z_1,0)&=&  \mathcal{I}_{T,L}^{LO}(\mathbf{x}_{0},\mathbf{x}_{1},1\!-\!z_1,z_1)\; \frac{{x}_{10}^2}{{x}_{20}^2\; {x}_{21}^2}
\label{ImpFact_NLO_soft_div}\, .
\end{eqnarray}

The general structure of the NLO dipole factorization formula \eqref{sigma_TL_Y_fplus} look similar to the one of the results of Balitsky and Chirilli for the high energy operator product expansion of two currents \cite{Balitsky:2010ze}. However, a more detailed comparison would require to perform a rather cumbersome Fourier transform of their result with respect to the positions of the two currents and the separation of the transverse and longitudinal photon contributions.

As expected, the evolution with respect to a change of the factorization scale $z_f$ (or equivalently $Y_f^+$) is governed by the B-JIMWLK equations. Indeed, the B-JIMWLK evolution of $\left\langle {\cal S}_{01} \right\rangle_{Y_f^+}$
writes
\begin{eqnarray}
\partial_{Y_f^+}  \left\langle {\cal S}_{01} \right\rangle_{Y_f^+}&=&   \bar{\alpha}
\int \frac{\textrm{d}^2\mathbf{x}_{2}}{2\pi} \frac{x_{01}^2}{x_{02}^2\, x_{21}^2}\: \left\langle{\cal S}_{02} {\cal S}_{21} \!-\! {\cal S}_{01} \right\rangle_{Y_f^+}
\, ,\label{B_JIMWLK_dipole}
\end{eqnarray}
and $\left\langle{\cal S}_{02} {\cal S}_{21} \!-\! {\cal S}_{01} \right\rangle_{Y_f^+}$ satisfies a similar but more complicated equation. Hence, the derivative with respect to $Y_f^+$ of the bracket in the second line of the equation \eqref{sigma_TL_Y_fplus} vanishes up to terms of order ${\cal O}(\bar{\alpha}^2)$.  The evolution of $\left\langle {\cal S}_{01}- {\cal S}_{02}\,{\cal S}_{21}\right\rangle_{Y_f^+}$ in the third line of the equation \eqref{sigma_TL_Y_fplus} also brings only irrelevant NNLO contributions. Finally, one might replace the lower bound $z_f$ of the $z_2$-integral of the third line by $0$, since the integrand is now regular.

In order to optimize the resummation of leading logs, one should obviously choose a factorization scale $z_f$ close enough to the upper part of its allowed range \eqref{z_range}. Accordingly, one should evolve $\left\langle {\cal S}_{01} \right\rangle_{Y_f^+}$ and $\left\langle{\cal S}_{02} {\cal S}_{21} \!-\! {\cal S}_{01} \right\rangle_{Y_f^+}$ with the B-JIMWLK equations over an interval $Y_f^+ \simeq \log\left((1\!-\!z_1)\, Q^2/x\, Q_0^2\right)$.

As a remark, notice that the dipole S-matrix $\left\langle {\cal S}_{01} \right\rangle_{Y_f^+}$ used in the factorization formula \eqref{sigma_TL_Y_fplus} or its LO approximation then depends on the Bjorken $x$ or $Q^2$ only through the combination $W^2= Q^2/x$, as advocated \emph{e.g.} in Refs.\cite{Forshaw:1999uf,Cvetic:1999fi}.


\section{Conclusion\label{sec:conclu}}

The main result presented here is the NLO generalization \eqref{sigma_TL_Y_fplus} of the dipole factorization formula for DIS structure functions at low $x$, with the longitudinal and transverse NLO impact factors \eqref{ImpFact_NLO_L} and \eqref{ImpFact_NLO_T}. In ref. \cite{Balitsky:2010ze}, results in principle equivalent have been already presented. However, in that paper, the results were given only in full position space for the incoming and outgoing photons, so that they were not ready for use in phenomenology before cumbersome Fourier transforms, by contrast to the results presented here. In future studies, it would be useful to perform those Fourier transforms, so that the independent NLO calculations of ref. \cite{Balitsky:2010ze} and of the present paper would cross-check each other.

Obviously, the results presented here pave the way towards phenomenological studies in the spirit of refs. \cite{Albacete:2009fh,Albacete:2010sy,Kuokkanen:2011je}, but at NLO accuracy. For consistency, however, one might have to use the present results together with \emph{e.g.} the NLL BK equation \cite{Balitsky:2008zz,Balitsky:2009xg}, whose numerical simulation seems to be a formidable task.

The calculations have been performed under the assumption that the target is dense, and thus include gluon saturation effects. However, in the 2 gluons exchange approximation, they should provide the appropriate NLO factorization formula to be used with the BFKL equation in mixed space.

As an intermediate step, the quark-antiquark-gluon components of the light front wave functions of virtual photons have been obtained, see equations \eqref{qqbarglue_wavefunction}, \eqref{NLO_T_wavefunction} and \eqref{NLO_L_wavefunction}. Those should be one of the building blocks for the NLO dipole factorization of other DIS observables, typically less inclusive.

With DIS phenomenology in mind, it would be useful to extend the results of the present study to quark mass effects. Indeed, the charm quark is known to provide a sizable contribution to DIS structure functions, and the bottom quark a non-negligible one. Quark mass effects are also required for phenomenology at low $Q^2$.

As a byproduct, the results presented here provide an interesting insight into the kinematics of initial state parton cascades in mixed space. It has been inferred that the prescription \eqref{trans_recoil} allows to take recoil effects exactly into account in mixed space. Finally, the generic formula \eqref{form_time_conjecture} for the mixed space expression for the formation time of a arbitrary $n$-partons state has been conjectured. That expression depends only on the final state of the shower, not on the precise diagram leading to that final state.


\begin{acknowledgments}
I thank Ian Balitsky and Giovanni Chirilli for explanations about their formalism and results.
I also acknowledge Yacine Mehtar-Tani, Al Mueller, Anna Sta\'sto and Bowen Xiao for useful comments on some of the results presented here.
Part of the work presented here has been performed under the Contract No. \#DE-AC02-98CH10886 with the
U.S. Department of Energy.
\end{acknowledgments}


\appendix


\section{Light-front perturbation theory for DIS at low $x$\label{sec:formalism}}

\subsection{Eikonal scattering and total cross section of a dilute projectile on a Color Glass Condensate}

Let us consider a right-moving relativistic projectile scattering on a very dense and highly boosted left-moving target. In DIS at low $x$ and similar processes, only the gluons with
low momentum fraction of the target are probed. For a dense enough target or at low enough $x$, the occupation number of those soft gluons becomes non-perturbatively large, so that the state of the target is more appropriately described in terms of semi-classical gluon fields than of Fock states with a finite number of gluons. This is the starting point \cite{McLerran:1993ni,McLerran:1993ka,McLerran:1994vd} of the Color Glass Condensate (CGC) effective theory  for the target and of related frameworks \cite{Balitsky:2001gj}.
For the collision of a dilute projectile on such a dense target, the disturbance of the semi-classical field of the target by the interaction is a higher order effect, so that only diagonal terms in the density matrix for the soft gluon field of the target are probed and thus the quantum average over the target gluon field reduces to a classical statistical average in the CGC formalism, up to perturbatively calculable corrections.

Then, one has to calculate the high-energy scattering of the projectile on an arbitrary classical gluon field. This can be done with the light-front formalism of ref.\cite{Bjorken:1970ah} generalized to QCD. In the limit of infinite boost, the target field-strength is Lorentz contracted to a shockwave localized at\footnote{In this study, the light-cone coordinates are defined as $x^\pm=(x^0\pm x^3)/\sqrt{2}$.} $x^+=0$. Hence, the interaction of the projectile with the classical field happens during an infinitely short $x^+$ time interval around $x^+=0$. In the Heisenberg picture, the projectile is thus in a state $|i_H\rangle$ for $x^+<0$ and switches to a state $\hat{S}_E |i_H\rangle$ for $x^+>0$, where $\hat{S}_E$ is a unitary operator which depends on the classical gluon field.

In the high-energy limit, the typical time-scale along $x^+$ for interactions between the partons in the projectile becomes infinitely larger than the $x^+$ time interval taken by the projectile to pass through the classical field, suggesting that these two types of interactions factorize from each other. Thus, different partons present in the projectile at $x^+=0$ should scatter independently from each other on the shockwave field.
The transverse position $\mathbf{x}$ of partons stays obviously constant during their infinitely fast interaction with the infinitely thin shockwave, but their transverse momentum $\mathbf{k}$ is typically modified. Moreover, in the limit of infinite boost of the target, the shockwave field is independent of $x^-$, so that it contains only gluon modes with $k^+=0$. Hence, the $k^+$ momentum of partons from the projectile cannot be modified by scattering on the shockwave field.
It is thus convenient to use a mixed space representation, specifying the longitudinal momentum $k^+$ and the transverse position $\mathbf{x}$ of partons in the projectile. We then use a Fock state basis to describe the state of the projectile at $x^+=0$, just before or just after the collision. The normalization of the annihilation operators $a$, $b$ and $d$ for gluons, quarks and anti-quarks is chosen such that
\begin{eqnarray}
\Big\{b(\mathbf{x}',{k'}^+,h',A',f') , b^\dag(\mathbf{x},{k}^+,h,A,f)   \Big\}&=& \Big\{d(\mathbf{x}',{k'}^+,h',A',f') , d^\dag(\mathbf{x},{k}^+,h,A,f)   \Big\}\nonumber\\
 &=& (2\pi)^3 2{k}^+\, \delta\big({k}^+\!-\!{k'}^+\big)\, \delta^{(2)}(\mathbf{x}\!-\!\mathbf{x}')\, \delta^{h,h'}\, \delta^{A,A'}\, \delta^{f,f'}\\
 \Big[a(\mathbf{x}',{k'}^+,\lambda',a') , a^\dag(\mathbf{x},{k}^+,\lambda,a)   \Big]&=&(2\pi)^3 2{k}^+\, \delta\big({k}^+\!-\!{k'}^+\big)\, \delta^{(2)}(\mathbf{x}\!-\!\mathbf{x}')\, \delta^{\lambda,\lambda'}\, \delta^{a,a'}\, ,
\end{eqnarray}
where $h=\pm 1/2$ and $\lambda=\pm 1$ are the helicities, $A$ and $a$ the fundamental and adjoint color indices, and $f$ the flavor. At high energy, the scattering of partons off the gluon shockwave is eikonal, so that $\hat{S}_E$ acts on Fock states by only color rotating each partons by a Wilson line defined along its trajectory through the shockwave, \emph{i.e.}
\begin{eqnarray}
& &\hat{S}_E\;  b^\dag(\mathbf{x}_l,k^+_l,h_l,A_l,f_l) \cdots d^\dag(\mathbf{x}_m,k^+_m,h_m,A_m,f_m)  \cdots a^\dag(\mathbf{x}_n,k^+_n,h_n,A_n,f_n) \cdots |0\rangle\nonumber\\
& & = \sum_{A_l, \dots} \sum_{A_m, \dots} \sum_{a_n, \dots} \left[U\left[{\cal A}\right](\mathbf{x}_l)\right]_{B_l {A_l}} \cdots \left[U^\dag\left[{\cal A}\right](\mathbf{x}_m)\right]_{{A_m} B_m} \cdots  \left[V\left[{\cal A}\right](\mathbf{x}_n)\right]_{b_n {a_n}}  \cdots\nonumber\\
& &\qquad\times\,  b^\dag(\mathbf{x}_l,k^+_l,h_l,{A_l},f_l) \cdots d^\dag(\mathbf{x}_m,k^+_m,h_m,{A_m},f_m)  \cdots a^\dag(\mathbf{x}_n,k^+_n,\lambda_n,{a_n}) \cdots |0\rangle\, ,\label{action_of_S_E}
\end{eqnarray}
where the fundamental and adjoint Wilson lines are respectively defined as the path-ordered exponential
\begin{eqnarray}
U\left[{\cal A}\right](\mathbf{x})&=& {\cal P}\, \exp \left[i g \int \textrm{d}x^+\, T^a\, {\cal A}^-_a(x^+,\mathbf{x},0) \right]\\
V\left[{\cal A}\right](\mathbf{x})&=& {\cal P}\, \exp \left[i g \int \textrm{d}x^+\, t^a\, {\cal A}^-_a(x^+,\mathbf{x},0) \right]
\end{eqnarray}
involving the classical shockwave gluon field ${\cal A}^\mu_a(x^+,\mathbf{x},x^-)$.

Finally, the projectile-shockwave total cross-section is obtained by the optical theorem, which writes in this case \cite{Bjorken:1970ah}
\begin{equation}
\sigma_{tot}[{\cal A}]= 2\; \frac{\textrm{Re}\left(\big\langle i_H({q'}^+)\big| 1\!-\! \hat{S}_E \big|i_H(q^+)\big\rangle\right)}{2 q^+\, (2\pi)\, \delta({q'}^+\!-\!q^+)}\, ,\label{OptTh}
\end{equation}
where $q^+$ is the momentum of the projectile.

\subsection{Light-front wave-function of the projectile in perturbation theory}

In order to calculate the total cross section of the projectile on a given shockwave field using eq.\eqref{OptTh}, we now only need to know how to write the incoming Heisenberg state of the projectile in terms of the Fock state basis for free on-shell partons at $x^+=0$. This can be done perturbatively in the framework of light-front wave functions \cite{Bjorken:1970ah}.
The light-front hamiltonian $\hat{{\cal P}}^-$ writes $\hat{{\cal P}}^-=\hat{{\cal T}}+\hat{{\cal U}}$, where $\hat{{\cal T}}$ is the free part and $\hat{{\cal U}}$ the interaction part (see ref.\cite{Brodsky:1997de} for more details). In the interaction picture, the $x^+$ evolution of operators is generated by $\hat{{\cal T}}$, for example
\begin{equation}
\hat{{\cal U}}_I(x^+)= e^{i \hat{{\cal T}} x^+}\, \hat{{\cal U}}_I(0)\, e^{-i \hat{{\cal T}} x^+}\, ,
\end{equation}
for the interaction operator, and the states $|i_I(x^+)\rangle$ evolve as
\begin{equation}
 \big|\, i_I(x_2^+)\big\rangle= {\cal P}\, \exp\left(  -i\int_{x_1^+}^{x_2^+} \textrm{d} x^+\, \hat{{\cal U}}_I(x^+)\right)\: \big|\, i_I(x_1^+)\big\rangle .\label{evol_state_I_pict}
\end{equation}
The interaction picture is defined here in such a way that it coincides at $x^+=0$ with the Heisenberg picture, \emph{i.e.} $|i_I(0)\rangle \equiv |i_H\rangle$. Then, expanding the exponential in \eqref{evol_state_I_pict} for $x_2^+=0$ and $x_1^+\rightarrow -\infty$ and inserting several times the decomposition of the identity
\begin{equation}
{ \mathbf{ 1}}=\sum_{\cal F} |{\cal F}\rangle \langle {\cal F}|,\label{identity_decomp_1}
\end{equation}
over a basis of Fock states $|{\cal F}\rangle$ at $x^+=0$, one finds
\begin{eqnarray}
|i_H\rangle &=& \sum_{{\cal F}_0}  \Big\langle {\cal F}_0 \Big|i_I(-\infty)\Big\rangle\, \Bigg\{|{\cal F}_0\rangle+ \sum_{n=1}^{\infty}  \sum_{{\cal F}_n}\cdots \sum_{{\cal F}_1} |{\cal F}_n\rangle\,\frac{1}{T_{{\cal F}_{0}}\!-\!T_{{\cal F}_{n}}+i\epsilon}  \, \langle {\cal F}_n| \hat{{\cal U}}_I(0)|{\cal F}_{n-1}\rangle \,\frac{1}{T_{{\cal F}_{0}}\!-\!T_{{\cal F}_{n\!-\!1}}+i\epsilon}  \, \cdots\nonumber\\
& &  \cdots  \,\frac{1}{T_{{\cal F}_{0}}\!-\!T_{{\cal F}_{1}}+i\epsilon}  \, \langle {\cal F}_1| \hat{{\cal U}}_I(0)|{\cal F}_0\rangle
\Bigg\}
\, ,\label{LFWF_pert}
\end{eqnarray}
where $T_{\cal F}$ is the eigenvalue of the free hamiltonian $\hat{{\cal T}}$ corresponding to the state $|{\cal F}\rangle$. Hence, we are forced to choose the standard momentum-space Fock basis instead of the mixed-space Fock basis considered previously, which does not diagonalize $\hat{{\cal T}}$. The two Fock bases are simply related by transverse Fourier transform of all the creation operators present in the state, with
\begin{equation}
b^\dag(\mathbf{k},k^+,h,A,f)=  \int \frac{\textrm{d}^2\mathbf{x}}{2\pi}\: e^{i \mathbf{k}\cdot\mathbf{x}}\;    b^\dag(\mathbf{x},k^+,h,A,f)\label{bdagFourier}
\end{equation}
and similar relations for $d^\dag$ and $a^\dag$.

Thus, the general method for the calculation is the following. First, calculate the momentum-space light-front wave function of the projectile $\langle {\cal F}|i_H\rangle$ to the appropriate order in perturbation theory using the general expression \eqref{LFWF_pert}. Then, perform the required Fourier transforms in order to get the light-front wave function in mixed-space. And finally, use the relations \eqref{action_of_S_E} and \eqref{OptTh} or analog ones, in order to obtain the total cross section or other observables for the scattering of the projectile off a given classical gluon field, before performing the CGC statistical average over the gluon field of the target.

Let us come back to the formula \eqref{LFWF_pert} in order to specify the last details. First, the eigenvalue $T_{\cal F}$ is simply the sum over the $k_i^-$ for the partons $i$ present in the Fock state. Remember that in light-front perturbation theory, only physical on-shell partons are included in the Hilbert space, so that $k_i^-={\mathbf{k}_i}^2/(2 k_i^+)$ for massless partons.

In the formal sum over Fock states used in equations \eqref{identity_decomp_1} and \eqref{LFWF_pert}, there is a summation over the number of partons of each type present in the Fock state.
And for each parton present, there is a summation over its quantum numbers and a phase-space integration
\begin{equation}
\int_{0}^{+\infty}\!\!\!\! \frac{\textrm{d}k^+}{(2\pi)(2 k^+)}\: \int \frac{\textrm{d}^2\mathbf{k}}{(2\pi)^2}\quad \dots
\end{equation}
consistent with the normalization of creation operators.

Inserting the expressions for the quantized free fields at $x^+=0$ into the interaction part of the hamiltonian $\hat{{\cal U}}_I(0)$, one can obtain the required expressions for the vertices.
In this study, we need two types of QCD vertices on the light-front: the quark to quark and gluon splitting and the antiquark to antiquark and gluon splitting, which write respectively
\begin{eqnarray}
\!\!\!\!\!\!\!&\!\!\!\!\!\!\! &\!\!\!\!\!\!\! \langle 0 | a(\mathbf{k}'',{k''}^+,\lambda,a) b(\mathbf{k}',{k'}^+,h',A',f')  \hat{{\cal U}}_I(0)  b^\dag(\mathbf{k},{k}^+,h,A,f)  |0\rangle  =  (2\pi)^3 \delta\big({k'}^+\!+\!{k''}^+\!-\!{k}^+\big)\: \delta^{(2)}\left(\mathbf{k}'\!+\!\mathbf{k}''\!-\!\mathbf{k}\right) \nonumber\\
& & \times\: \delta_{f, f'}\: \delta_{h, h'}\: g\, \big(T^a\big)_{A' A}\, \sqrt{4 {k}^+ {k'}^+}\: \varepsilon_{\lambda}^{*}\cdot \Bigg[ \frac{\mathbf{k}''}{{k''}^+}\!-\!\left(\frac{1\!+\!(2h)\lambda}{2}\right)\frac{{\mathbf{k}'}}{{k'}^+}
\!-\!\left(\frac{1\!-\!(2h)\lambda}{2}\right)\frac{\mathbf{k}}{{k}^+} \Bigg]\label{q2qg_vertex}
\end{eqnarray}
and
\begin{eqnarray}
\!\!\!\!\!\!\!&\!\!\!\!\!\!\! &\!\!\!\!\!\!\! \langle 0 | a(\mathbf{k}'',{k''}^+,\lambda,a) d(\mathbf{k}',{k'}^+,h',A',f')  \hat{{\cal U}}_I(0)  d^\dag(\mathbf{k},{k}^+,h,A,f)  |0\rangle  =  (2\pi)^3 \delta\big({k'}^+\!+\!{k''}^+\!-\!{k}^+\big)\: \delta^{(2)}\left(\mathbf{k}'\!+\!\mathbf{k}''\!-\!\mathbf{k}\right) \nonumber\\
& & \times\: (-1) \delta_{f, f'}\: \delta_{h, h'}\: g\, \big(T^a\big)_{A A'}\, \sqrt{4 {k}^+ {k'}^+}\: \varepsilon_{\lambda}^{*}\cdot \Bigg[ \frac{\mathbf{k}''}{{k''}^+}\!-\!\left(\frac{1\!+\!(2h)\lambda}{2}\right)\frac{{\mathbf{k}'}}{{k'}^+}
\!-\!\left(\frac{1\!-\!(2h)\lambda}{2}\right)\frac{\mathbf{k}}{{k}^+} \Bigg]\label{qbar2qbarg_vertex}
\end{eqnarray}
in the case of massless quarks. Here, $\varepsilon_{\lambda}$ is the transverse polarization vector for transverse gauge bosons of helicity $\lambda$, \emph{i.e.}
\begin{equation}
\varepsilon_{\lambda}=\frac{1}{\sqrt{2}}\: \left(
                                              \begin{array}{c}
                                                1 \\
                                                i \lambda \\
                                              \end{array}
                                            \right)
\, ,
\end{equation}
which satisfies the relations
\begin{equation}
\sum_{\lambda=\pm 1} \varepsilon_{\lambda}^{i *}\;  \varepsilon_{\lambda}^{j} = \delta^{ij} \qquad \textrm{and} \qquad
\sum_{\lambda=\pm 1} \lambda\: \varepsilon_{\lambda}^{i *}\;  \varepsilon_{\lambda}^{j} = i\: \epsilon^{ij}
\end{equation}
and $\epsilon^{ij}$ is antisymmetric with $\epsilon^{12}=+1$.

We also need the QED vertex for the splitting of a photon into a quark and antiquark dipole
\begin{eqnarray}
\!\!\!\!\!\!\!&\!\!\!\!\!\!\! &\!\!\!\!\!\!\! \langle 0 | d(\mathbf{k}_1,k^+_1,h_1,A_1,f_1)  b(\mathbf{k}_0,k^+_0,h_0,A_0,f_0)  \hat{{\cal U}}_I(0)  {a_{\gamma}}^\dag(\mathbf{k},k^+,\lambda)   |0\rangle  =  (2\pi)^3 \delta\big({k_0}^+\!+\!{k_1}^+\!-\!{k}^+\big)\: \delta^{(2)}\left(\mathbf{k}_0\!+\!\mathbf{k}_1\!-\!\mathbf{k}\right) \nonumber\\
& & \times\:  e\, e_{f_0}\: \delta_{f_0, f_1}\: \delta_{A_0, A_1}\: \delta_{h_0, -h_1}\, \sqrt{4 k^+_0 k^+_1}\: \varepsilon_{\lambda}\cdot \Bigg[ \frac{\mathbf{k}}{{k}^+}\!-\!\left(\frac{1\!+\!(2h_0)\lambda}{2}\right)\frac{\mathbf{k}_1}{k_1^+}
\!-\!\left(\frac{1\!-\!(2h_0)\lambda}{2}\right)\frac{\mathbf{k}_0}{{k}_0^+} \Bigg]\, ,\label{gamma2qqbar_vertex}
\end{eqnarray}
and the mixed QED/QCD vertex for the instantaneous splitting of a photon into a quark, an antiquark and a gluon
\begin{eqnarray}
\!\!\!\!\!\!\!&\!\!\!\!\!\!\! &\!\!\!\!\!\!\! \langle 0 | a(\mathbf{k}_2,k^+_2,\lambda_2,a) d(\mathbf{k}_1,k^+_1,h_1,A_1,f_1)  b(\mathbf{k}_0,k^+_0,h_0,A_0,f_0)  \hat{{\cal U}}_I(0)  {a_{\gamma}}^\dag(\mathbf{k},k^+,\lambda)   |0\rangle =  (2\pi)^3 \delta\big({k_0}^+\!+\!{k_1}^+\!+\!{k_2}^+\!-\!{k}^+\big)  \nonumber\\
& & \times\, \delta^{(2)}\left(\mathbf{k}_0\!+\!\mathbf{k}_1\!+\!\mathbf{k}_2\!-\!\mathbf{k}\right)\:  e\, e_{f_0}\: \delta_{f_0, f_1}\: g\, \big(T^a\big)_{A_0 A_1}\: \delta_{h_0, -h_1}\: \delta_{\lambda, \lambda_2}\, \sqrt{4 k^+_0 k^+_1}\:  \Bigg[\frac{\delta_{\lambda, -2 h_0}}{k^+\!-\!k^+_1}\!-\!\frac{\delta_{\lambda, 2 h_0}}{k^+\!-\!k^+_0}\Bigg]\, ,\label{inst_gamma2qqbarg_vertex}
\end{eqnarray}
where $e$ is the proton electric charge and $e_f$ the fractional charge of the quark of flavor $f$.

\subsection{DIS case: from lepton to virtual photon scattering\label{sec:lepton2photon}}

In DIS, the initial asymptotic state of the projectile $|i_I(-\infty)\rangle$ contains only one lepton, and $|{\cal F}_0\rangle=|i_I(-\infty)\rangle$. As usual, only the LO contribution to DIS with respect to QED is relevant: the initial lepton couples to quarks via a single photon exchange. To that order, one can trivially factor out the leptonic part of the diagram in covariant perturbation theory, and then one has to study the scattering off the target of the intermediate virtual photon which can be transverse or longitudinal. In our formalism based on light-front perturbation theory, there are neither virtual particles nor longitudinal photons in the Hilbert space, so that the factorization of the leptonic tensor is much less obvious but still holds, for the following reasons.

In all energy denominators of the formula \eqref{LFWF_pert}, the momentum $(k_l^+,\mathbf{k}_l)$ of the incoming lepton appears via $T_{{\cal F}_{0}}=k_l^-=\mathbf{k}_l^2/(2 k_l^+)$. And the scattered lepton is present in all of the Fock states corresponding to intermediate or final steps in \eqref{LFWF_pert}, in addition to the emitted photon or to colored partons. Hence the second term $T_{{\cal F}_{m}}$ (with $m>0$) in all the energy denominators contains the contribution   ${k_l'}^-={\mathbf{k}_l'}^2/(2 {k_l'}^+)$ of the scattered lepton. Because of momentum conservation, the momentum of the emitted photon is $q^+=k_l^+\!-\!{k_l'}^+$ and $\mathbf{q}=\mathbf{k}_l\!-\!\mathbf{k}_l'$. Although it cannot be interpreted as photon virtuality in this light-front formalism, it is still convenient to introduce the variable
\begin{equation}
Q^2=- \big(k_l^\mu\!-\!{k_l'}^\mu\big)\big({k_l}_\mu\!-\!{k_l'}_\mu\big)\, .
\end{equation}
In a frame in which $\mathbf{q}=0$, the contribution of the incoming and scattered leptons to any energy denominator writes
\begin{equation}
k_l^-\!-\!{k_l'}^-=- \frac{Q^2}{2 q^+}\, .
\end{equation}

Apart from that, the incoming and scattered leptons are involved in the formula \eqref{LFWF_pert} only in the first interaction vertex. Actually, DIS processes can be initiated in two ways in the present formalism: either the lepton emits a transverse photon, which then propagates and splits later into a quark antiquark dipole, or the quark antiquark pair production occurs in one step via instantaneous coulombian interaction between the quark and leptonic currents. The latter case can be understood as involving an instantaneous longitudinal photon exchange, so that this photon is not present in any of the intermediate Fock states ${\cal F}_{m}$. Thanks to the integration over the azimuthal angle of the outgoing lepton performed for DIS, the interference between those transverse and longitudinal contributions vanish, so that they add up incoherently in the DIS cross section.

The leptonic part of the contribution with transverse photon can be factorized out easily: just remove the first step in \eqref{LFWF_pert}, including the lepton to lepton and photon vertex and the energy denominator for the lepton-photon intermediate Fock state. Then, one can use all the formalism presented before, but using the transverse photon instead of the lepton as projectile and replacing $T_{{\cal F}_{0}}$ by $-Q^2/(2 q^+)$ in \eqref{LFWF_pert}, and calculate the total cross section $\sigma^{\gamma}_T$ for the scattering of the transverse photon off the target.

By contrast, for the longitudinal contribution, one has to separate by hand the lepton to lepton quark and anti-quark vertex into a factor associated to the QED coupling of the leptons, a factor analog to an energy denominator, and a factor associated to the QED pair production of quarks. This last factor is interpreted as a fictitious vertex for the splitting of a longitudinal virtual photon into a quark anti-quark dipole, and writes
\begin{equation}
{\cal V}_{\gamma^*_L(q^+,Q^2)\rightarrow q(\mathbf{k},k^+,h,A,f)\; \bar{q}(\mathbf{k}',{k'}^+,h',A',f') }  =  (2\pi)^3 \delta\big({k}^+\!+\!{k'}^+\!-\!{q}^+\big)\: \delta^{(2)}\left(\mathbf{k}\!+\!\mathbf{k}'\right)\: \delta_{f, f'}\: \delta_{h', -h}\: \delta_{A, A'}\: e\: e_f\, \sqrt{4 k^+ {k'}^+}\: \frac{Q}{q^+}\, .\label{gammaL2qqbarVertexEff}
\end{equation}
Then, the total cross section $\sigma^{\gamma}_L$ for the scattering of the longitudinal photon off the target is calculated using this splitting vertex instead of the transverse one \eqref{gamma2qqbar_vertex} and the replacement of $T_{{\cal F}_{0}}$ by $-Q^2/(2 q^+)$ in energy denominators.

Hence, the hadronic part of the diagrams relevant to DIS can be calculated separately from the leptonic part in light-front perturbation theory as well as in covariant perturbation theory, leading to the photon target cross sections $\sigma^{\gamma}_T$ and $\sigma^{\gamma}_L$. And the full DIS cross section in the one photon exchange approximation is obtained via the same expression as in covariant perturbation theory
\begin{equation}
\frac{\textrm{d}^2\, \sigma^{DIS}}{ \textrm{d}x\, \textrm{d} Q^2}= \frac{\alpha_{em}}{\pi\, x\, Q^2} \left\{\left(1\!-\!y\!+\!\frac{y^2}{2}\right)\, \sigma^{\gamma}_T(x,Q^2)+ \left(1\!-\!y\right)\, \sigma^{\gamma}_L(x,Q^2)  \right\}
\, .
\end{equation}
The variable $y$ is defined as
\begin{equation}
y=\frac{Q^2}{x s}\, .
\end{equation}
$s$ is the Mandelstam $s$ variable for the lepton-target collision, and the Bjorken $x$ variable can be approximated for our purposes as
\begin{equation}
x=\frac{Q^2}{2 (q\cdot P)}\simeq \frac{Q^2}{2 q^+\, P^-}\, .
\end{equation}
The energy-momentum momentum of the target $P^\mu$ has a big component $P^-$, whereas $\mathbf{P}=0$ by choice of frame, and $P^+$ is given by the on-shellness condition $M_t^2=P^\mu P_\mu=2 P^-\, P^+$.

Naively, $\sigma^{\gamma}_T(x,Q^2)$ and $\sigma^{\gamma}_L(x,Q^2)$ look independant of the Bjorken $x$ in our formalism. Indeed, $|i_H\rangle$ describes the projectile independently of the target, the eikonal scattering operator $\hat{S}_E$ seems independent of the collision energy, as long as it is large enough for the formalism to be valid, and the CGC average over the gluon field of the target seems independent of the projectile.
However, the contributions to $|i_H\rangle$ containing gluons at $x^+=0$ have the usual divergence of \emph{Bremsstrahlung} in the soft gluon limit. And the Wilson lines have rapidity divergences for generic field of the target. These two types divergences can be regulated by a common cut-off in longitudinal momentum or rapidity, as explained in section \ref{sec:subtract_LL}. Then, one has a renormalization group evolution when moving this cut-off, which is the B-JIMWLK evolution in general, and reduces to the BK equation in a mean field approximation and to the BFKL equation in the dilute target limit. Then $\sigma^{\gamma}_T(x,Q^2)$ and $\sigma^{\gamma}_L(x,Q^2)$ become dependent on $x$ through the factorization scale associated to that evolution.


\section{A few integrals}

When calculating the wave-functions for transverse and longitudinal virtual photons in mixed space up to NLO, one encounters the following integrals:

\begin{eqnarray}
\int\frac{\textrm{d}^2\mathbf{k}}{2\pi}\, \frac{e^{i \mathbf{k}\cdot \mathbf{x}}}{\left.\overline{Q}^2+\mathbf{k}^2\right.}\; \mathbf{k}^j&=& i\, \frac{\mathbf{x}^j}{|\mathbf{x}|}\:  \overline{Q}\; \textrm{K}_1\!\left(\overline{Q}\, |\mathbf{x}| \right)\label{K1_LO}\\
\int\frac{\textrm{d}^2\mathbf{k}}{2\pi}\, \frac{e^{i \mathbf{k}\cdot \mathbf{x}}}{\left.\overline{Q}^2+\mathbf{k}^2\right.}&=& \textrm{K}_0\!\left(\overline{Q}\, |\mathbf{x}| \right)\label{K0_LO}\\
\int\frac{\textrm{d}^2\mathbf{k}_1}{2\pi}\, \int\frac{\textrm{d}^2\mathbf{k}_2}{2\pi}\,  \frac{e^{i \mathbf{k}_1\cdot \mathbf{x}_{10}+i \mathbf{k}_2\cdot\mathbf{x}_{20}} \quad \mathbf{k}_1^j \;  \left(\frac{\mathbf{k}_{2}^m}{z_2}\!+\!\frac{\mathbf{k}_{1}^m}{1\!-\!z_1} \right)}{
\left(z_1(1\!-\!z_1)Q^2+{\mathbf{k}_1}^2\right)\left(Q^2\!+\!\frac{(\mathbf{k}_1+\mathbf{k}_2)^2}{(1\!-\!z_1\!-\!z_2)}\!
+\!\frac{\mathbf{k}_1^2}{z_1}\!+\!\frac{\mathbf{k}_2^2}{z_2}\right)}& &\nonumber\\
 =-z_1(1\!-\!z_1\!-\!z_2)\, \left(\mathbf{x}_{10}^j\!-\!\frac{z_2}{1\!-\!z_1} \mathbf{x}_{20}^j\right) \frac{\mathbf{x}_{20}^m}{{x}_{20}^2}\:  \frac{QX\; \textrm{K}_1\!\left(QX\right)}{X^2}& &\label{K1_NLO_2steps}\\
\int\frac{\textrm{d}^2\mathbf{k}_1}{2\pi}\, \int\frac{\textrm{d}^2\mathbf{k}_2}{2\pi}\,  \frac{e^{i \mathbf{k}_1\cdot \mathbf{x}_{10}+i \mathbf{k}_2\cdot\mathbf{x}_{20}}}{
\left(Q^2\!+\!\frac{(\mathbf{k}_1+\mathbf{k}_2)^2}{(1\!-\!z_1\!-\!z_2)}\!
+\!\frac{\mathbf{k}_1^2}{z_1}\!+\!\frac{\mathbf{k}_2^2}{z_2}\right)}&=& z_2\, z_1(1\!-\!z_1\!-\!z_2)\,   \frac{QX\; \textrm{K}_1\!\left(QX\right)}{X^2}\label{K1_NLO_inst}\\
\int\frac{\textrm{d}^2\mathbf{k}_1}{2\pi}\, \int\frac{\textrm{d}^2\mathbf{k}_2}{2\pi}\,  \frac{e^{i \mathbf{k}_1\cdot \mathbf{x}_{10}+i \mathbf{k}_2\cdot\mathbf{x}_{20}} \quad   \left(\frac{\mathbf{k}_{2}^m}{z_2}\!+\!\frac{\mathbf{k}_{1}^m}{1\!-\!z_1} \right)}{
\left(z_1(1\!-\!z_1)Q^2+{\mathbf{k}_1}^2\right)\left(Q^2\!+\!\frac{(\mathbf{k}_1+\mathbf{k}_2)^2}{(1\!-\!z_1\!-\!z_2)}\!
+\!\frac{\mathbf{k}_1^2}{z_1}\!+\!\frac{\mathbf{k}_2^2}{z_2}\right)}&=& i\,\frac{(1\!-\!z_1\!-\!z_2)}{(1\!-\!z_1)}\, \frac{\mathbf{x}_{20}^m}{{x}_{20}^2}\:  \textrm{K}_0\!\left(QX\right)\, .\label{K0_NLO_2steps}
\end{eqnarray}
Here, $X$ is defined by Eq.\eqref{def_X_a}. All those integrals can be performed by using the Schwinger representation of the denominators in order to transform the integrals over the transverse momentums into gaussian ones.


\section{Formation time for $4$-particle states\label{sec:4part_form_time}}

\begin{figure}
\setbox1\hbox to 10cm{
\fcolorbox{white}{white}{
  \begin{picture}(402,252) (9,1)
    \SetWidth{1.4}
    \SetColor{Black}
    \Photon(10,132)(90,132){3.5}{4}
    \Line(160,152)(300,232)
    \Line[dash,dashsize=10](160,152)(300,192)
    \SetWidth{1.0}
    \Line[dash,dashsize=2](300,242)(300,32)
    \Text(280,12)[lb]{\Large{\Black{$x^+=0$}}}
    \Text(310,192)[lb]{\Large{\Black{$\mathbf{x}_{0'}$}}}
    \Text(310,232)[lb]{\Large{\Black{$\mathbf{x}_{0}$}}}
    \Text(310,162)[lb]{\Large{\Black{$\mathbf{x}_{3}$}}}
    \Text(221,196)[lb]{\Large{\Black{$z_{0}$}}}
    \Text(90,152)[lb]{\Large{\Black{$z_{0}\!+\!z_2\!+\!z_3$}}}
    \SetWidth{1.4}
    \Line(230,142)(300,162)
    \Text(310,62)[lb]{\Large{\Black{$\mathbf{x}_{1}$}}}
    \Text(160,92)[lb]{\Large{\Black{$z_{1}$}}}
    \Text(100,32)[lb]{\Huge{\Black{$(A)$}}}
    \Line(90,132)(300,72)
    \Line(90,132)(160,152)
    \Text(170,132)[lb]{\Large{\Black{$z_{2}\!+\!z_3$}}}
    \Line(160,152)(230,142)
    \Line[dash,dashsize=10](230,142)(300,132)
    \Text(310,122)[lb]{\Large{\Black{$\mathbf{x}_{2'}$}}}
    \Text(310,92)[lb]{\Large{\Black{$\mathbf{x}_{2}$}}}
    \Line(230,142)(300,102)
    \Text(240,112)[lb]{\Large{\Black{$z_{2}$}}}
    \Text(240,152)[lb]{\Large{\Black{$z_{3}$}}}
    \SetWidth{1.0}
    \SetColor{White}
    \EBox(10,2)(410,242)
  \end{picture}
}
}
\setbox2\hbox to 10cm{
\fcolorbox{white}{white}{
  \begin{picture}(402,242) (9,-9)
    \SetWidth{1.0}
    \SetColor{White}
    \EBox(10,-8)(410,232)
    \SetWidth{1.4}
    \SetColor{Black}
    \Photon(10,122)(90,122){3.5}{4}
    \Line(90,122)(190,152)
    \Line(190,152)(290,202)
    \Line(190,92)(290,92)
    \Line[dash,dashsize=10](190,152)(290,182)
    \SetWidth{1.0}
    \Line[dash,dashsize=2](290,232)(290,22)
    \Text(270,2)[lb]{\Large{\Black{$x^+=0$}}}
    \Text(300,177)[lb]{\Large{\Black{$\mathbf{x}_{0'}$}}}
    \Text(300,197)[lb]{\Large{\Black{$\mathbf{x}_{0}$}}}
    \Text(300,142)[lb]{\Large{\Black{$\mathbf{x}_{2}$}}}
    \Text(300,57)[lb]{\Large{\Black{$\mathbf{x}_{1'}$}}}
    \Text(221,186)[lb]{\Large{\Black{$z_{0}$}}}
    \Text(220,132)[lb]{\Large{\Black{$z_{2}$}}}
    \Text(220,102)[lb]{\Large{\Black{$z_{3}$}}}
    \Text(90,142)[lb]{\Large{\Black{$z_{0}\!+\!z_2$}}}
    \SetWidth{1.4}
    \Line(90,122)(190,92)
    \Line[dash,dashsize=10](190,92)(290,62)
    \Text(90,92)[lb]{\Large{\Black{$z_{1}\!+\! z_3$}}}
    \Line(190,92)(290,42)
    \Line(190,152)(290,152)
    \Text(300,32)[lb]{\Large{\Black{$\mathbf{x}_{1}$}}}
    \Text(300,82)[lb]{\Large{\Black{$\mathbf{x}_{3}$}}}
    \Text(200,52)[lb]{\Large{\Black{$z_{1}$}}}
    \Text(100,22)[lb]{\Huge{\Black{$(B)$}}}
  \end{picture}
}
}
\begin{center}
\resizebox*{12cm}{!}{\hspace{-2cm}\mbox{\box1 \hspace{4cm} \box2}}
\caption{\label{Fig:Topo4partons}The two possible topologies (A) and (B) for diagrams without instantaneous interactions leading to a $4$-partons Fock state.}
\end{center}
\end{figure}
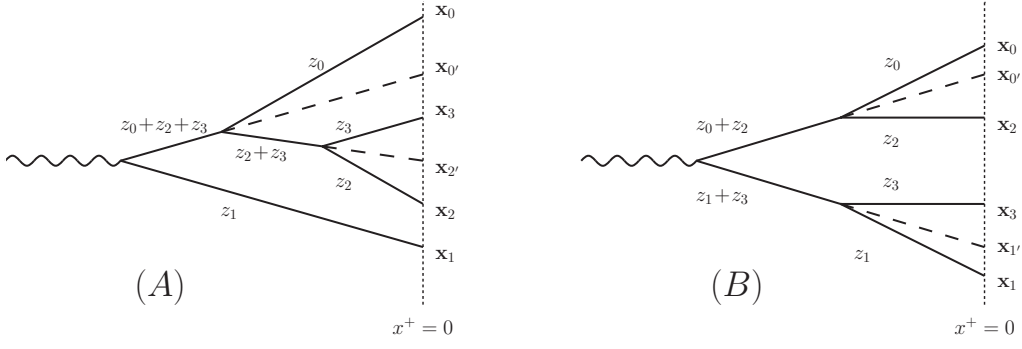

Excluding diagrams with instantaneous interactions, there are two generic topologies for tree level diagrams describing the splitting of one particle into four, up to permutations of particles. Those two topologies are illustrated in Fig.\ref{Fig:Topo4partons}. Following the prescription \eqref{trans_recoil} for recoil effects, one writes in the case of the topology (A)
\begin{eqnarray}
\mathbf{x}_{2'}&=&\frac{z_2\, \mathbf{x}_{2}+ z_3\, \mathbf{x}_{3}}{z_2\!+\!z_3}\\
\mathbf{x}_{0'}&=&\frac{z_0\, \mathbf{x}_{0}+ (z_2\!+\!z_3)\, \mathbf{x}_{2'}}{z_0\!+\!z_2\!+\!z_2}=\frac{z_0\, \mathbf{x}_{0}+ z_2\, \mathbf{x}_{2}+ z_3\, \mathbf{x}_{3}}{z_0\!+\!z_2\!+\!z_2}\, .
\end{eqnarray}
Then, one constructs the variable $X_{4,\, (A)}^2$ by summing the formation times in mixed space associated with each of the three vertices
\begin{eqnarray}
X_{4,\, (A)}^2&=&z_1\, (z_0\!+\!z_2\!+\!z_3)\, x_{0'1}^2+\frac{z_0\, (z_2\!+\!z_3)}{z_0\!+\!z_2\!+\!z_3}\, {x}_{02'}^2+\frac{z_2\, z_3}{z_2\!+\!z_3}\, {x}_{23}^2\nonumber\\
&=&z_1\, (z_0\!+\!z_2\!+\!z_3)\,
\left(\frac{z_0\, \mathbf{x}_{0}+ z_2\, \mathbf{x}_{2}+ z_3\, \mathbf{x}_{3}}{z_0\!+\!z_2\!+\!z_2}-\mathbf{x}_{1} \right)^2\nonumber\\
& &+\frac{z_0\, (z_2\!+\!z_3)}{z_0\!+\!z_2\!+\!z_3}\, \left(\mathbf{x}_{0}- \frac{z_2\, \mathbf{x}_{2}+ z_3\, \mathbf{x}_{3}}{z_2\!+\!z_3}\right)^2
+\frac{z_2\, z_3}{z_2\!+\!z_3}\, {x}_{23}^2\, .
\end{eqnarray}

In an analogous way, one has for the topology (B)
\begin{eqnarray}
\mathbf{x}_{0'}&=&\frac{z_0\, \mathbf{x}_{0}+ z_2\, \mathbf{x}_{2}}{z_0\!+\!z_2}\\
\mathbf{x}_{1'}&=&\frac{z_1\, \mathbf{x}_{1}+ z_3\, \mathbf{x}_{3}}{z_1\!+\!z_3}\, .
\end{eqnarray}
And the variable $X_{4,\, (B)}^2$ sums the formation time for the three splitting present for that topology
\begin{eqnarray}
X_{4,\, (B)}^2&=&(z_0\!+\!z_2)(z_1\!+\!z_3)\, x_{0'1'}^2+\frac{z_0\, z_2}{z_0\!+\!z_2}\, {x}_{02}^2+\frac{z_1\, z_3}{z_1\!+\!z_3}\, {x}_{13}^2\nonumber\\
&=&(z_0\!+\!z_2)(z_1\!+\!z_3)\, \left(\frac{z_0\, \mathbf{x}_{0}+ z_2\, \mathbf{x}_{2}}{z_0\!+\!z_2}\!-\!
\frac{z_1\, \mathbf{x}_{1}+ z_3\, \mathbf{x}_{3}}{z_1\!+\!z_3} \right)^2+\frac{z_0\, z_2}{z_0\!+\!z_2}\, {x}_{02}^2+\frac{z_1\, z_3}{z_1\!+\!z_3}\, {x}_{13}^2\, .
\end{eqnarray}

It is then possible, using the constraint $z_0\!+\!z_1\!+\!z_2\!+\!z_3=1$, to show that both $X_{4,\, (A)} ^2$ and $X_{4,\, (B)} ^2$ reduce to the expression $X_{4} ^2$ given by the equation \eqref{form_time_conjecture}
\begin{equation}
X_{4,\, (A)} ^2=X_{4,\, (B)} ^2 =X_{4} ^2 = z_0\, z_1\, {x}_{01}^2+z_0\, z_2\, {x}_{02}^2+z_0\, z_3\, {x}_{03}^2
+z_1\, z_2\, {x}_{12}^2+z_1\, z_3\, {x}_{13}^2+z_2\, z_3\, {x}_{23}^2
\, .\label{form_time_4part}
\end{equation}
Hence, even for larger $n$, the variable $X_{n} ^2$ associated to the formation time of a $n$-partons state seems still independent on the path taken by the parton cascade. The diagrams with instantaneous interactions should not destroy that property, like in the $3$ partons case. Indeed, due to gauge invariance, they should not feature so different kinematic properties than non-instantaneous diagrams.


\bibliography{MaBiblioHEQCD}


\end{document}